\documentclass[11pt,draftclsnofoot,onecolumn]{IEEEtran}

\usepackage{amsmath}
\input{amssym}
\usepackage{graphicx}
\usepackage{caption}
\usepackage{subcaption}

\ifCLASSINFOpdf
\else
\fi
\begin{document}
%
\title{Bare Demo of IEEEtran.cls for Journals}
%
%
%

\author{Nasser~Eslahi,~\IEEEmembership{Student~Member,~IEEE,}
                Ali~Aghagolzadeh,~\IEEEmembership{Senior~Member,~IEEE}
and~Seyed~Mehdi~Hosseini~Andargoli,~\IEEEmembership{Member,~IEEE}
\thanks{This paper was presented in part the 7th International Symposium on Telecommunications, Tehran, Iran, Sep. 2014.}
\thanks{The authors are with the Faculty of Electrical and Computer Engineering, Babol University of Technology, Babol~47148-71167,~Iran (e-mail: nasser.eslahi@stu.nit.ac.ir; aghagol@nit.ac.ir; smh$\_$andargoli@nit.ac.ir)}}

%
%

\markboth{Draft Paper}%
{Shell \MakeLowercase{\textit{et al.}}:Compressive Video Sensing via Adaptive Dictionary Learning}
%



\title{Compressive Video Sensing via Dictionary Learning and Forward Prediction}

\maketitle

\begin{abstract}
In this paper, we propose a new framework for compressive video sensing (CVS) that exploits the inherent spatial and temporal redundancies of a video sequence, effectively. The proposed method splits the video sequence into the key and non-key frames followed by dividing each frame into the small non-overlapping blocks of equal size. At the decoder side, the key frames are reconstructed using adaptively learned sparsifying (ALS) basis via $\ell_0$ minimization, in order to exploit the spatial redundancy. Also, three well-known dictionary learning algorithms are investigated in our method. For the recovery of non-key frames, a prediction of current frame is initialized, by using the previous reconstructed frame, in order to exploit the temporal redundancy. The prediction is employed into a proper optimization problem to recover the current non-key frame. To compare our experimental results with the results of some other methods, we employ peak signal to noise ratio (PSNR) and structural similarity (SSIM) index as the quality assessor. The numerical results show the adequacy of our proposed method in CVS.
\end{abstract}


%
\IEEEpeerreviewmaketitle

\section{Introduction}
%
%
%
%
\IEEEPARstart{C}{onventional} approaches to image/video compression usually have high computational complexity in encoding but they remain simple in the decoding process part. Also, distributed video coding (DVC) [1] is a low complexity technique which refers to a special video coding paradigm that encodes frames of video sequence independently (the encoder can be very simple) and decodes them jointly at the expense of a more complex decoder (due to exploitation of the temporal redundancies by the decoder side). In both conventional video coding and DVC, data collection and compression tasks are performed disjointedly, with a high cost mechanism that wastes most valuable acquired data, because of limitation of allocated power and available memory. 
Due to great efforts by Cand\`{e}s \emph{et al}. [2], [3] and Donoho [4], compressive sensing (also called compressed sensing or compressive sampling) suggests a new framework for simultaneous sampling and compression of signals at a rate significantly below the Nyquist rate. It also permits that under certain conditions, the original signal can be reconstructed properly from a small set of measurements via solving a convex optimization problem or iterative greedy recovery algorithms.

Recently, the idea of compressive sensing (CS) for imaging (single pixel camera [5, 6]) has been extended to the conventional predictive/distributed video coding, to develop highly desirable compressive video sensing (CVS)/distributed compressive video sensing (DCVS). CVS employs both data acquiring (video sensing) and compression into a unified task which emerges a new procedure to directly acquiring compressed video data via random projection for each individual frame, in a low-complexity encoder.

\subsection{Related Works}Several CVS recovery methods have already been proposed, i.e., Wakin \emph{et al}. [7] proposed an intuitive (motion JPEG motivated) approach which extends compressive image sensing to video applications by considering each frame of the video sequence independently, and recovers each frame using the 2D discrete cosine transform (2D DCT) or a 2D discrete wavelet transform (2D DWT), individually. However, this simple extension fails to address the temporal redundancy in video; nevertheless compressed image sensing techniques explore the spatial redundancy within an image. To enhance the signal sparsity in both spatial and temporal domains and achieve higher sampling efficiency, several frames can be jointly considered as a signal and recovered under a 3D transform (e.g., 3D DWT) [7, 8]. Results show that a 3D video reconstruction (joint frames) using a 3D DWT is better than the 2D frame-by-frame reconstruction using a 2D DWT, but it incurs high computation cost and large memory requirement.

Park and Wakin [9] proposed a multi-scale recovery approach, which several CS measurements are taken independently for each frame, and also the motion estimation is applied at the decoding step. The recovered video at coarse scales (low spatial resolution) is used to estimate motion which is then used to enhance the recovery at finer scales (high spatial resolutions). However, its coarse-to-fine reconstruction computational cost is high. The same approach based on using two-steps to iteratively update the estimates for the images in the video and the inter frame motion, was proposed in [10]. Also, Cossalter \emph{et al}. [11] considered the motion estimation in their proposed joint compressive video coding and analysis scheme.

Stankovi\'{c} \emph{et al}. [12] and Prades-Nobet \emph{et al}. [13] proposed a block-based selective video sampling scheme where firstly divides frames of the video sequence into reference frames (or key frames) and non-reference frames (or non-key frames); then each frame is divided into the small non-overlapping blocks of equal size. The reference frames are sampled fully using conventional video compression techniques like MPEG/H.264 (intra encoding). In [12], the reference frames are used to predict sparsity of the successive non-reference frames and CS is applied only within the blocks that are predicted as sparse ones, whereas the remaining blocks are sampled fully. In [13], non-key frames are projected and recovered using CS techniques, with an adaptive redundant dictionary built by picking a set of local (spatially neighboring) blocks which are extracted from its co-located blocks in the previous decoded key frame, to form its basis without training. A similar approach was proposed in [14]. Nevertheless, such local dictionary-based basis may not work very well for blocks with large motion or when the entire scene undergoes motion translation. In addition such schemes highly rely on the qualities of the neighboring reconstructed key frames and the performance may be degraded due to poorly reconstructed neighboring key frames [15]. 

The work in [16] relies on small inter-frame differences together with a spatial 2D DWT to produce a sparse representation of the underlying video. Similarly, Zheng and Jakobs [17] proposed a video compressive sensing method that relies on the sparsity properties of video in the spatial domain, where key (reference) frames are fully sampled and measurements are applied to the difference between each pair of successive frames. The video signal is reconstructed by first reconstructing the frame differences using $\ell_{1}$ minimization algorithm, and then adding them sequentially to the reference frame. However, they do not sufficiently remove the vast amount of temporal redundancy, e.g., when exist a large inter-frame difference and fast motion between adjacent frames.

A multi-hypothesis (MH) prediction approach for CVS was proposed in [18], where MH predictions of the current frame are generated from one or more previously reconstructed reference frames, and then combined to yield a composite prediction superior to any of the constituent single-hypothesis predictions.

Another dictionary based approach is presented in [19], where the dictionary is learned from a set of blocks globally extracted from the previous reconstructed neighboring frames together with the side information generated from them. This work has been extended to assign dynamic measurement rate allocation (for different local regions) by incorporating a feedback channel [15]. Haixao \emph{et al}. [20] proposed a redundant dictionary generation scheme for compressed video sensing, which follows the sparse representation approach of [13]-[14]. The authors extended their work to present a maximum-likelihood dictionary learning based reconstruction algorithm for DCVS [21]. Up to our best knowledge, there also exist other literatures about CS-based video coding, e.g., [22]-[23].

\subsection{Contribution of This Paper}
In this paper we propose a novel for sampling and recovering of compressed sensed video data. The proposed method divides the video sequence into the key and non-key frames follow by dividing each frame into the small blocks of equal sizes, similar to [13]. Each blocks of key (non-key) frames are sampled using the same sensing matrix $\Phi_{{B}_{K}}$ ($\Phi_{{B}_{NK}}$). The compressed key frame data are reconstructed initially using MH block compressed sensing recovery [18], in order to use as the initial image in the dictionary leaning algorithm step, to obtain an adaptively learned sparsifying (ALS) basis to exploit the spatial redundancy of frame, by an iterative procedure. Also, we investigate the effectiveness of three well-known dictionary learning methods, to adopt in our scheme. The obtained ALS basis is incorporated into the optimization problem, for the whole CS frame recovery, in form of $\ell_{0}$ quasi-norm. In this step, a split Bregman iteration (SBI) [24]-[25] based technique is employed to solve the non-convex $\ell_{0}$ minimization, efficiently. For the recovery of non-key frames, first we initialize a prediction of current frame, $\hat{F}_{t}$, by means of previous reconstructed frame $\tilde{F}_{t-1}$, in order to exploit the temporal redundancy. The prediction $\hat{F}_{t}$ is employed into a SBI based method together with the achieved ALS basis of current frame and current CS data, to recover the current non-key frame $\tilde{F}_{t}$ by solving the proper minimization problem. The experimental results show the high competitive performance of our proposed method compared with the other state-of-the-art CVS techniques.

The rest of this paper is organized as follows. In Section II, we introduce an abstract framework for CS theory and SBI method. Also, we review three well-known techniques of dictionary learning, briefly. The proposed method is described in Section III. Numerical results and comparisons for our proposed method are given in Section IV and finally, Section V concludes the paper.
\section{Background}
This section reviews an abstract framework of CS theory and SBI algorithm, accompanied by investigation of three well-known dictionary learning methods, briefly.
\subsection{Compressed Sensing}
Suppose we wish to recover a real value finite length signal $u\in \Bbb{R}^n$ from a finite length observation $f \in \Bbb{R}^m$; so that $ m\ll n$ and there is a linear projection between them $$f_{m \times 1} = \Phi_{m \times n} u_{n \times 1}+\emph{e}_{{m \times 1}} \eqno (1)$$ where $\Phi \in \Bbb{R}^{m \times n}$  is a sensing matrix and $\emph{e} \in \Bbb{R}^{m \times 1}$ denotes the additive noise. Since the number of unknowns is much more than the observations, clearly we are not able to recover every $u$ from $f$ and it is generally considered as an \emph{ill-posed} problem. However, if $u$ be sufficiently sparse in the sense that $u$ can be written as a superposition of a small number of vectors taken from a known (sparsifying) transform domain basis $(t=n)$ or frame $(t>n)$ $\Psi \in \Bbb{R}^{t \times n}$ or even adaptive learned sparsifying basis (e.g., see [26]), then exact recovery of $u$ is possible. So sparsity plays a key role in recovering of $u$ from observation vector $f$. Also, $ u$ would be called $s$-sparse if only its $s$  coefficients in the set of transform domain ($\vartheta = \Psi u$) are nonzero and the other $n-s$ coefficients are zero. In other words, the transform domain signal $\vartheta$ can be well approximated using only $s < m\ll n$ nonzero entries. In order to solve the reconstruction problem with a reasonable accuracy and robustness to the noise, the estimation of $u$ is formulated as an unconstrained Lagrangian optimization problem which incorporates the prior information about the original signal
$$\mathop{\min}_{u} \{{1 \over 2}\|f-\Phi u\|_{\ell_2}^2 +\lambda\|\Psi u\|_{\ell_p}\} \eqno(2)$$
or
$$\mathop{\min}_{\vartheta} \{{1 \over 2}\|f-\Phi\Psi^{-1}\vartheta\|_{\ell_2}^2 +\lambda\|\vartheta\|_{\ell_p}\}, \eqno(3)$$
where $u=\Psi^{-1} \vartheta$.
Here, the first term is a penalty that represents the closeness of the solution to the observed scene and quantifies the “prediction error” with respect to the measurements. The second term is a regularization term that represents a priori sparse information of the original scene and also it is designed to penalize an estimate that would not exhibit the expected properties. Also, $\lambda$ is a regularization parameter that balances the contribution of both terms. In the second term of (2) or (3), $\ell_{p}$ is usually considered as $\ell_{0}$ or $\ell_{1}$. This minimizing problem can be solved easily by iterative shrinkage/thresholding (IST) methods (see, e.g., [27]), Bregman iterative algorithms (see, e.g.,~[24] and [25]). Since $\ell_{0}$ minimization is non-convex and its solution is considered as \emph{NP-hard}, the common method is to replace $\ell_{0}$ quasi-norm with the $\ell_{1}$ norm, because it is the closest convex norm to non-convex $\ell_{0}$ quasi-norm and minimizing the $\ell_{1}$ norm instead. However, a fact that is often neglected is, for some practical problems, i.e., image inverse problems, the conditions guaranteeing the equivalence of $\ell_{0}$ minimization and $\ell_{1}$ minimization are not necessarily satisfied.
\subsection{Split Bregman Iteration}
The split Bregman iteration (SBI) method was recently proposed by Goldstein and Osher [24] for effectively solving $\ell_{1}$-regularized optimization problem with multiple $\ell_{1}$-regularized terms [25]. The basic idea of SBI is to convert the unconstrained minimization problem into a constrained one by introducing the variable splitting technique and then invoke the Bregman iteration [28] to solve the constrained minimization problem. The rationale behind SBI is that it may be easier to solve the constrained problem than to solve its unconstrained counterpart. Another advantage of the SBI is that it has relatively small memory footprint and is easy to program by users. Such properties are very attractive for large-scale problems.

Consider an unconstrained optimization problem$$\mathop{min}_{v \in \Bbb{R}^{N}}\{F_{1}(v)+F_{2}(Gv)\} \eqno(4)$$where $G \in \Bbb{R}^{M \times N}$, $F_{1}:\Bbb{R}^{N}\rightarrow \Bbb{R}$ and $F_{2}:\Bbb{R}^{M}\rightarrow \Bbb{R}$. Apparently, (4) can be converted into an equivalent constrained form and then solved by invoking the SBI. The SBI framework is presented in Table I.
\begin{table}[!t]
\renewcommand{\arraystretch}{1}
{\footnotesize{{~~~~~~~~~~~~~~~~~TABLE I: The SBI A\sc lgorithm Framework}}}\\\\
\label{table_example}
\centering
\begin{tabular}{l}
\hline\hline
 {\bfseries{Input}}: $\{F_{1}(v)\}$,$\{F_{2}(w)\}$, $G$, $\mu>0$ \\
 {\bfseries{Output}}: $v$: Reconstructed signal;\\
 {\bfseries{Initialization}}: Set $k=0$; $(v^{0},w^{0},b^{0})=(0,0,0)$\\ 
 {\bfseries{While}} a stop criterion is not satisfied  {\bfseries{do}}\\
1.~~~~~$v^{k+1}=\arg\min_{v}{{F_{1}}(v)+{\frac{\mu}{2}}{\|w^{k}-Gv-b^{k}\|_{\ell_{2}}^2}}$\\
2.~~~~~$w^{k+1}=\arg\min_{w}{{F_{2}}(w)+{\frac{\mu}{2}}{\|w-Gv^{k+1}-b^{k}\|_{\ell_{2}}^2}}$\\
3.~~~~~$b^{k+1}=b^{k}+Gv^{k+1}-w^{k+1}$\\
4.~~~~~$k\leftarrow k+1$\\
{\bfseries{End}} While\\
\hline
{\bfseries{Note}}: $w=Gv$.\\ \hline 
\end{tabular}
\end{table}
\subsection{Sparse Representation and Dictionary Learning}
As stated previously, the key of the sparse representation modeling lies in the choice of sparsifying basis (or dictionary D). In this context, $D=[d_{1}, d_{2}, \cdots, d_{t}]\in \Bbb{R}^{n \times t}$ is called a dictionary and each of its columns is called an atom. One crucial problem in a sparse-representation problem is how to choose the efficient dictionary. There are many pre-specified (non-adaptive analytically designed) sparsifying dictionaries (basis or frame), e.g., Fourier, discrete cosine transform, wavelets, Ridgelets, Curvelets, Contourlets, Shearlets and etc. In spite of being simple and having fast computations, the analytically designed dictionaries are not able to efficiently (sparsely) represent a given class of signals, and they lack the adaptivity to the image local structures. However, learning the atoms from a set of training signals belonging to signal class of interest would result in dictionaries with the capability of better matching the content of the signals [29]. It has been experimentally shown that these adaptive dictionaries outperform the non-adaptive ones in many signal processing applications.

Dictionary learning algorithms iteratively perform the two stages of \emph{sparse approximation} (\emph{sparse coding}) and \emph{dictionary update}. In the first stage, which is actually the clustering of the signals into a union subspace, the sparse approximation of the signals is computed using the current dictionary. The second stage is the update of the dictionary. To the best of our knowledge, most dictionary learning algorithms differ mainly in the way of updating the dictionaries [30]-[32]. Some algorithms such as K-singular value decomposition (K-SVD) [31] are based on updating the atoms one-by-one, while some others such as method of optimal directions (MOD) [30] updates the whole set of atoms at once. In [32], a MOD-like algorithm was proposed in which more than one atom along with the non-zero entries in their associated row vectors in coefficient matrix are updated at a time. We refer to this algorithm as the multiple dictionary update (MDU) algorithm (for further reading see [32]).
\subsection{Patch-based redundant sparse recovery}
In literature, the basic unit of sparse representation for natural images is patch [31]. Suppose the vector representation of the original image denotes by $u\in \Bbb{R}^n$, accordingly, $u_{{p}_{l}}\in \Bbb{R}^{B_s}$ represents an image patch of size ${\sqrt{B_s}}\times {\sqrt{B_s}}$ at location $l$, $l=1, 2, \cdots, J$. Then we have$$u_{{p}_{l}}=R_{l} u, \eqno(5)$$where $R_{l}\in \Bbb{R}^{B_{s}\times {n}}$ is a binary matrix operator that extracts the square patch $u_{l}$ from $u$, forming the output patch as the column vector. Note that, patches are usually overlapped (to suppress the boundary artifacts), and such patch based representation is highly redundant and significant to achieve high recovery quality. Therefore, the recovery of $u$ from $\{u_{{p}_{l}}\}$ becomes an over-determind system, which is straightforward  to obtain the following least-square solution:
$$u={({\sum_{l=1}^{J}R_{l}^{T}R_{l}})^{-1}}{{\sum_{l=1}^{J}}(R_{l}^{T}u_{{p}_{l}})}, \eqno(6)$$
which is nothing but telling that the overall image is reconstructed by averaging all the overlapped patches. Given dictionary $D$, the sparse coding process of each patch $u_{{p}_{l}}$ over $D$ is to find a sparse vector $\alpha_{l}$ such that $u_{{p}_{l}}\approx D\alpha_{l}$. Then the entire image can be sparsely represented by the set of sparse codes $\{\alpha_{l}\}$:
$$u\approx D\circ\alpha={({\sum_{l=1}^{J}R_{l}^{T}R_{l}})^{-1}}{{\sum_{l=1}^{J}}(R_{l}^{T}D\alpha_l)}, \eqno(7)$$
where $\alpha$ denotes the concatenation of all $\{\alpha_l\}$.
By given a set of training image patches $P=[u_{{p}_{1}}, u_{{p}_{2}}, \cdots, u_{{p}_{{J}'}}]$, where $J'$ is the number of training image patches, the goal of sparsifying basis learning is to jointly optimize the sparsifying basis $D$, and the representation coefficient matrix $\Lambda=[\alpha_{1}, \alpha_{2}, \cdots, \alpha_{J'}]$; such that $u_{{p}_{l}}=D \alpha_{l}$ and $\|\alpha_{l}\|_{\ell_{p}}\leq L$, where $\ell_{p}$ is $\ell_{0}$ or $\ell_{1}$. This can be formulated by the following minimization problem:
$$(\hat{D}, \hat{\Lambda})={\arg\min_{D, \Lambda}}\sum_{l=1}^{J'}{\|u_{{p}_{l}}-D\alpha_{l}\|_{\ell_{2}}^2}~~s.t.~~\|\alpha_{l}\|_{\ell_{p}}\leq L, \forall{l}, \eqno(8)$$ where the requirement of $\|\alpha_l\|_{\ell_0}\leq L\ll n$ indicates that the sparse representation stage uses no more than $L$ atoms from the dictionary for every image patch instance.

Although the above minimization problem in equation (8) is large-scale and highly non-convex even when $\ell_{p}$ is equal to $\ell_{1}$, some approximation approaches (e.g., MOD [30], K-SVD [31] and MDU [32]) have been proposed to optimize $D$ and $\Lambda$ alternatively, leading to many state-of-the-art results in image processing.
\section{The Proposed Method}
By contrast with the conventional/distributed video coding scheme, in which data acquiring and compression tasks are performed disjointedly, CVS employs both data acquiring (video sensing) and compression, into a unified task which emerges a new procedure to directly acquiring compressed video data via random projection (without temporally storing the complete raw data) for each individual frame (or blocks of frame) at a low complexity encoder. In this case, the majority of computational burden is shifted from the encoder to the decoder side, which is more suitable to deploy in modern video applications, e.g., video serveillance systems and wireless multimedia sensor networks. In this section, first we discuss about encoding of the proposed method, and then we propose decoding scheme of key and non-key frames.
\subsection{Encoding}
As mentioned before, the proposed method, firstly, divides the video sequence into the key and non-key frames follow by dividing each frame, of size $I_{r}\times I_{c}$, into small non-overlapping blocks of equal size (i.e., size $B\times B$), and then the same sensing (sampling/measurement) matrix $\Phi_{B}${\footnote{Note that, $m\over n$ denotes the measurement ratio; the measurement ratio for key and non-key frame may differ. In this context, $MR_{K}$ ($MR_{NK}$) and ${\Phi_{B}}_{K}$ (${\Phi_{B}}_{NK}$) are used as the measurement ratio and the sensing matrix of key (non-key) frames, respectively.}} (i.e., size $m_{B}\times B^2$, where $m_{B}=\lfloor{mB^2\over n}\rfloor, n=I_{r} I_{c}$) is applied for sampling of eah block. In this case, we have:$$f_{i}=\Phi_{B} u_{i} \eqno(9)$$where $u_{i}$ is a (column) vector representing of block $i$ of the input image, $f_{i}$ is its corresponding measurement vector and $\Phi_{B}$ independently samples blocks within frame. Using this technique has several benefits comparing to use of a random sampling operator to the entire image; i.e., the encoder does not need to wait until the entire image is measured, but each block is sent after its linear projection. In addition, at the decoder side, each block is processed independently; therefore the speed of encoding and decoding procedure is increased. Also in this case, we just need to store a $m_{B}\times B^2$ sensing matrix instead of a $m\times n$ sensing matrix. More precisely, the global sensing matrix takes a block-diagonal structure, $\Phi={\rm diag}(\Phi_{B},\cdots,\Phi_{B})$.
\subsection{Recovery of key frame}
At the decoder side, the key frame is reconstructed initially using method of the multi-hypothesis block CS recovery [11], in order to use as initial training image in the process of learning an ALS basis (dictionary). Now, in this case, by considering $u\approx D\circ\alpha$, The equation (3) using ALS basis can be written as$$\mathop{\min }_{\alpha} \{{1 \over 2}\|f-\Phi D\circ\alpha\|_{\ell_2}^2 +\lambda\|\alpha\|_{\ell_p}\}. \eqno(10)$$

Here, $D$ replaces $\Psi^{-1}$ in equation (3), standing for ALS basis. Also, $\alpha$ denotes the patch-based redundant sparse representation for the whole image over $D$, which can be find by solving (10). As discussed previously, when $\ell_{p}$ is replaced with $\ell_{0}$, since $\ell_{0}$ is non-convex and NP-hard, the usual routine is to solve its optimal convex approximation, i.e., $\ell_{1}$ minimization. However, for some practical problems, i.e., image inverse problems, the conditions guaranteeing the equivalance of $\ell_{0}$ minimization and $\ell_{1}$ minimization are not necessarily satisfied. An approach was proposed in [26], where equation (10) can be efficiently solved via ALS basis and $\ell_{0}$ minimization. In this part, we adopt the proposed scheme of [26] to solve (10), which leads to recovery of the key frames. 
Now, let`s go back to equation (10) and point out how to solve it. By considering $v=D\circ\alpha$ and $\ell_{p}=\ell_{0}$ (sparsity is strictly measured), (10) can be formulated into an equivalent constrained form as:
$$\mathop{\min}_{\alpha, v} \{{1\over 2}\|f-\Phi v\|_{\ell_2}^2 +\lambda\|\alpha\|_{\ell_0}\}~~s.t.~~v=D\circ\alpha. \eqno(11)$$

In order to solve the above minimization problem, an alternating SBI algorithm (as illustrated in Table. 1) is applied. We finally obtain the following schemes:
$$v^{k+1}=\arg\min_{v}{{1\over 2}{\|f-\Phi v\|_{\ell_2}^2}+{\frac{\mu}{2}}{\|D\circ\alpha^{k}-v-b^{k}\|_{\ell_{2}}^2}}, \eqno(12)$$
$$\alpha^{k+1}=\arg\min_{\alpha}{\lambda}\|\alpha\|_{\ell_0}+{\frac{\mu}{2}}{{\|D\circ\alpha-v^{k+1}-b^{k}\|_{\ell_{2}}^2}}, \eqno(13)$$
$$b^{k+1}=b^{k}+v^{k+1}-D\circ\alpha^{k+1}. \eqno(14)$$ 
Here, $\mu$ is a fixed parameter for improving the numerical stability of the algorithm.

Given $\alpha^{k}$, the sub-problem of (12) consists in minimizing a strictly convex quadratic function, that can be solved easily. By setting the gradient of the objective function in (12) to be zero, a closed solution for (12) is achieved (see the Appendix I), which can be expressed as $$v^{k+1}={(\Phi^{T}\Phi+\mu I)}^{-1}{\big(\mu(D\circ\alpha^{k}-b^{k})+\Phi^{T}f\big)}, \eqno(15)$$where $(\cdot)^{-1}$ and $I$ denote the matrix inverse operator and the identity matrix, respectively. Since for image CS recovery and also here, $\Phi$ is a random projection matrix. Therefore, it is too costly to solve the minimization of the quadratic function in (12) directly by using (15), because of existence of the matrix inverse in (15). Here, in order to avoid computing the inversion of matrix, the steepest descent method with the optimal step is utilized to solve the minimization of the quadratic function in (12), which can be expressed as$$v^{k+1}=v^{k}-\eta^{k} g^{k}. \eqno(16)$$
Here, $g$ is the gradient direction of the objective function and $\eta$ represents the optimal step size; for finding $\eta$, see the Appendix I.

Now, by given $v$ in hand (according to (12)), and by considering $r=v+b$ and $u=D\circ\alpha$ (for simplicity, the subscript $k$ is dropped without confusion), the sub-problem of (13) becomes
$$\min_{\alpha}{{1\over 2}{\|u-r\|_{\ell_{2}}^{2}+{\lambda \over \mu}{\|\alpha\|_{\ell_{0}}}}}. \eqno(17)$$
 
By these transformations, we regard $r$ as some type of the noisy observation of $u$. However, it is worth to note that, due to the complicated definition of $\alpha$, it is difficult to solve (17) directly. In order to solve (17) amenable, in this paper, a reasonable assumption is used, which leads to obtain a closed-form solution of (17). Such this assumption is efficiently employed in [14] and [33], hence we utilize the same approach to solve the above minimization problem. 

Let denote the error vector by $e=u-r$ (such that $u,r\in \Bbb{R}^{n}$) and each element of $e$ by $e(i), i=1,\cdots,n$. Also, we suppose that each element of $e$ $(e(i))$ follows an independent zero-mean distribution, $E\{e(i)\}=0$, with the same variance $Var\{e(i)\}=E\{e^{2}(i)\}=\sigma^{2}$, where $Var\{\cdot\}, E\{\cdot\}$ represent the variance and the expectation operator, respectively. It is worth emphasizing that the above assumption does not need to be Gaussian, or Laplacian, or generalized Gaussian distribution process, which are more general. By invoking the \emph{law of large numbers} in probability theory, for any $\epsilon>0$, it leads to $\lim_{n\rightarrow\infty}\Pr\{|{{1}\over{n}}\sum_{i=1}^{n}e^{2}(i)-\sigma^{2}|<{{\epsilon}\over{2}}\}=1$, i.e., 
$$\lim_{n\rightarrow\infty}\Pr\{{\big|}{{1}\over{n}}\|u-r\|_{\ell_{2}}^{2}-\sigma^{2}{\big|}<{{\epsilon}\over{2}}\}=1, \eqno(18)$$
where $\Pr\{\cdot\}$ represents the probability.

Let $u_{c}, r_{c}$ denote the concatenation of all the patches $u_{{p}_{l}}$ and $r_{{p}_{l}}$, $l=1,2,\cdots,J$, respectively, and each element of $u_c-r_c$ is denoted by $e_c(j), j=1,\cdots,K$, where $K=B_s\times J$. In accordance with the assumption, it is concluded that $e_c(j)$ is independent with zero mean and variance $\sigma^{2}$. Thus, due to the \emph{law of the large numbers} and by doing the same manipulation with (18) to $e_c^{2}(j)$, it yields $\lim_{K\rightarrow\infty}\Pr\{|{{1}\over{K}}\sum_{j=1}^{K}e_{c}^{2}(j)-\sigma^{2}|<{{\epsilon}\over{2}}\}=1$, i.e., 
$$\lim_{K\rightarrow\infty}\Pr\{{\big|}{{1}\over{K}}\sum_{l=1}^{J}\|u_{{p}_{l}}-r_{{p}_{l}}\|_{\ell_{2}}^{2}-\sigma^{2}{\big|}<{{\epsilon}\over{2}}\}=1. \eqno(19)$$

Therefore, according to (18) and (19), the following property is concluded
$$\lim_{\substack{{n\rightarrow\infty}\\{K\rightarrow\infty}}}\Pr\{{\big|}{{1}\over{n}}\|u-r\|_{\ell_{2}}^{2}-{{1}\over{K}}\sum_{l=1}^{J}\|u_{{p}_{l}}-r_{{p}_{l}}\|_{\ell_{2}}^{2}{\big|}<{{\epsilon}}\}=1, \eqno(20)$$
in which, the relationship between $\|u-r\|_{\ell_2}^2$ and $\sum_{l=1}^{J}\|u_{{p}_{l}}-r_{{p}_{l}}\|_{\ell_{2}}^{2}$ (with large probability) is described:
$${{1}\over{n}}\|u-r\|_{\ell_{2}}^{2}={{1}\over{K}}\sum_{l=1}^{J}\|u_{{p}_{l}}-r_{{p}_{l}}\|_{\ell_{2}}^{2}. \eqno(21)$$
Now, by substituting (21) into (17), $J$ sub-problems for all the patches $u_{{p}_{l}}$ is achieved that they can be solved more efficiently. Each patch based sub-problem is formulated as
$$\arg\min_{\alpha_l}{{1}\over{2}}\|u_{{p}_{l}}-r_{{p}_{l}}\|_{\ell_{2}}^{2}+\theta\|\alpha_l\|_{\ell_0}, \eqno(22)$$
where $\theta={{\lambda K} \over {\mu n}}$. In fact, by considering $u_{{p}_{l}}=D\alpha_l$, where $D$ is the adaptive learned dictionary from $r_{{p}_{l}}$ using patch-based dictionary learning method described in Section II-D, the above sub-problem can be considered as the sparse coding problem, i.e.,
$$\arg\min_{\alpha_l}{{1}\over{2}}\|D\alpha_{l}-r_{{p}_{l}}\|_{\ell_{2}}^{2}+\theta\|\alpha_l\|_{\ell_0}. \eqno(23)$$
Also, in order to achieve higher sparsity, (23) can be formulated in its constrained form, i.e.,
$$\min_{\alpha_l}\|\alpha_l\|_{\ell_0}~~~s.t.~~~\|D\alpha_{l}-r_{{p}_{l}}\|_{\ell_{2}}^{2}\leq \delta \eqno(24)$$
where $\delta=\omega \theta$ is a small constant controlling the approximation error, and $\omega$ is a control factor. Now, (24) can be solved efficiently using orthogonal matching pursuit (OMP) [34] algorithm. However, if $\ell_0$ pseudo-norm in (24) was relaxed with $\ell_1$ convex norm, it could be solved with basis pursuit [35], lasso [36] and $\ell_1$-regularized least square [37], but may be at the cost of less sparse solution. For all $J$ overlapped patchs, this process is employed to achieve ${\alpha}$ (concatenation of all $\{\alpha_l\}, l=1,\cdots,J$), which is the solution of the sub-problem (17).

Thus, the key frame is recovered efficiently, by solving the optimization problem of (11) using discussed SBI method (e.g., see (12)-(14)) via updating dictionary using (8) in which, $u_{{p}_{l}}$ is replaced by $r_{l}$, (since $r$ is regarded as a good approximation of $u$ at each iteration, here, we conduct adaptive sparsifying basis learning using all the patches extracted from $r$),$$\hat{D}={\arg\min_{D}}\sum_{l=1}^{J'}{\|r_{{p}_{l}}-D\alpha_{l}\|_{\ell_{2}}^2}~~s.t.~~\|\alpha_{l}\|_{\ell_{0}}\leq L, \forall{l}. \eqno(25)$$
The proposed algorithm for recovery of key frame data is described in detail in Table II.
\begin{table}[!t]
\renewcommand{\arraystretch}{1}
{\footnotesize{{TABLE II: T\sc he Detailed Description of Key Frame Recovery Framework}}}\\\\
\label{table_example}
\centering
\begin{tabular}{l}
\hline\hline
 {\bfseries{Input}}: $f$, $\Phi_B, B_s, \omega, \lambda, iin$: inloop iteration number\\ 
~~~~~~~~~$k_{max}$: maximum iteration number, $Tol$: Tolerance\\
 {\bfseries{Output}}: $u^*=D\circ\alpha^{*}$: Recovered key frame;\\
 {\bfseries{Initialization}}: Set $k=0$; $(\alpha^{0},b^{0})=(0,0)$\\
$u_{init}=v^{0}$= MH\_recovery($f$,$\Phi_B$) using method of [18] \\
 {\bfseries{While}} a stop criterion is not satisfied  {\bfseries{do}}\\
1.~~~~~$v^{k+1}=\tilde{v}=v^{k}$\\
~~~~~~~$r^{k+1}=v^{k+1}+b^{k}$\\
2.~~~~~update $D^{k+1}$ using (25)\\
3.~~~~~{\bfseries{for}} each patch $u_{p_{l}}$ {\bfseries{do}}\\
~~~~~~~~~~compute $\alpha_{l}^{k+1}$ using (24)\\
~~~~~~~{\bfseries{end}} for\\
4.~~~~~update $\alpha^{k+1}$ by concatenating all $\{\alpha_{{l}}^{k+1}\}$\\
5.~~~~~{\bfseries{for}} $i=1:iin$\\
~~~~~~~~~~$g_{i}^{k} \leftarrow \Phi^{T}\Phi v^{k}-\Phi^{T}f-\mu(D^{k+1}\circ\alpha^{k+1}-v^{k+1}-b^{k})$\\
~~~~~~~~~~$\eta_{i}^{k} \leftarrow {\rm diag}\big({\rm abs}({{{g_{i}^{k}}^{T}g_{i}^{k}}\over {{g_{i}^{k}}^{T}(\Phi^{T}\Phi+\mu I)g_{i}^{k}}})\big)$\\
~~~~~~~~~~$v^{k+1} \leftarrow v^{k+1}-{\eta}_{i}^{k}g_{i}^{k}$\\  
~~~~~~~{\bfseries{end}} for\\
6.~~~~~compute $s^{k+1}$= SSIM($v^{k+1}, \tilde{v}$)\\
7.~~~~~compute \emph{diff} = {\rm {abs}}$(s^{k+1}-s^{k})$\\
8.~~~~~update $b^{k}$ using (14)\\
9.~~~~~$k\leftarrow k+1$\\
{\bfseries{end}} While\\
\hline
{\bfseries{stopping criterion}}: $k=k_{max}$ or {\emph{diff}} $\leq Tol$$$.\\ \hline 
\end{tabular}
\end{table}
\subsection{Recovery of non-key frame} 
While spatial domain compression is performed by CS, the temporal redundancy is not exploited fully, because no motion estimation and compensation is performed at the CVS encoder. To incorporate the temporal redundancy, in order to efficient recovery of the non-key frames, the temporal correlation between adjacent frames is exploited through the inter-frame sparsity model. Here, an iterative approach for the reconstruction of the non-key frames is adopted, where the approach initially estimates an approximation of the non-key frame using previous reconstructed frame, in order to take the advantage of the inherent temporal structure between successive frames. Then, the initially estimated frame is utilized into an optimization problem to recover and refine the current non-key frame. 

Assume that $u_{t-1}^{*}=D\circ\alpha_{t-1}^{*}$, where $u_{t-1}^{*}$ is the previous reconstructed frame using ALS basis $D$, and $\alpha_{t-1}^{*}$ denotes the patch-based redundant sparse representation for the whole previous reconstructed frame over $D$. The initialization step can be formulated as:
$$\mathop{\min }_{\alpha} \{{1 \over 2}\|f-\Phi D\circ\alpha\|_{\ell_2}^2 +\lambda\|\alpha\|_{\ell_1}+\tau{\|\alpha-\alpha_{t-1}^{*}}\|_{\ell_1}\}, \eqno(26)$$
where the first term keeps the solution close to the measurements, the second term promotes sparsity in the spatial transform of the current frame, and the third term promotes sparsity in the inter-frame difference to achieve the temporal redundancy between the current frame and the previous reconstructed frame. In this paper, the problem of (26)  is solved efficiently by an alternating SBI-based framework via an iterative-shrinkage method using surrogate function.

Assume that $v=D\circ\alpha$, then (26) can be formulated in its constrained form:
$$\mathop{\min}_{\alpha, v} \{{1 \over 2}\|f-\Phi v\|_{\ell_2}^2 +\lambda\|\alpha\|_{\ell_1}+\tau{\|\alpha-\alpha_{t-1}^{*}}\|_{\ell_1}\}~~s.t.~~v=D\circ\alpha. \eqno(27)$$ 
Now, the above equation can be solved easily by an alternating SBI-based framework as follow: 
\begin{equation}
v^{k+1}=\arg\min_{v}\big\{{1 \over 2}\|f^{k}-\Phi v\|_{\ell_2}^2+{\mu\over 2}\|D\circ\alpha^{k}-v-b^{k}\|_{\ell_{2}}^{2}\big\}, \nonumber\\
\end{equation}
\begin{equation}
\alpha^{k+1}=\arg\min_{\alpha}\big\{{\frac{\mu}{2}}{{\|D\circ\alpha-v^{k+1}-b^{k}\|_{\ell_{2}}^2}}+{\lambda}\|\alpha\|_{\ell_1}+\tau\|\alpha-\alpha_{t-1}^{*}\|_{\ell_1}\big\}, \nonumber\\
\end{equation}
\begin{equation}
b^{k+1}=b^{k}+v^{k+1}-D\circ\alpha^{k+1}, \nonumber\\
\end{equation}
$$
f^{k+1}=f^{k}+f-\Phi v^{k+1}. \eqno(28)
$$
The last term in (28), adds a residual feedback to the algorithm.

Given $\alpha^{k}$, $b^{k}$ and $f^{k}$, the first sub-problem in (28) is essentially a minimization problem of strictly convex quadratic function, $Q_{2}(v)$, that is 
$$\min Q_{2}(v)=\min_{v}\big\{{1 \over 2}\|f^{k}-\Phi v\|_{\ell_2}^2+{\mu\over 2}\|D\circ\alpha^{k}-v-b^{k}\|_{\ell_{2}}^{2}\big\}. \eqno(29)$$
Setting the gradient of the objective function in (29) to be zero, leads to a closed solution resemble to that expressed in (15), but, here   $f$ is replaced with $f^{k}$.
Resemble (15), here exists matrix inverse that is too costly to compute, when $\Phi$ is a random projection matrix. Therefore, again the steepest descent method is used to solve (29), efficiently.

By given $v^{k+1}$, and assuming $u=D\circ\alpha$ and $r=v+b$ (for the simplicity the subscript $k$ is omitted without confusion), the $\alpha$ sub-problem in (28) can be formulated as
$$\min_{\alpha}\big\{{1\over 2}{\|u-r\|_{\ell_{2}}^{2}+{\lambda \over \mu}{\|\alpha\|_{\ell_{1}}}+{\tau\over \mu}{\|\alpha-\alpha_{t-1}^{*}}\|_{\ell_1}}\big\}. \eqno(31)$$
Now, by utilizing the same assumption which used in Section III-B, the sub-problem of (31) can be formulated as follow
$$\min_{\alpha} \sum_{l=1}^{J} \big\{{1 \over 2}\|u_{p_{l}}-r_{p_{l}}\|_{\ell_2}^{2}+{\lambda K \over \mu n}\|\alpha_l\|_{\ell_1}+{\tau K \over \mu n}\|\alpha_l-\alpha_{t-l_{l}}^{*}\|_{\ell_1}\big\}. \eqno(32)$$
Obviously, (32) can be solved efficiently by solving $J$ sub-problems for all patches. By assuming $u_{p_{l}}=D\alpha_{l}$, $\theta_{1}={{\lambda K}\over{\mu n}}$, $\theta_{2}={{\tau K}\over{\mu n}}$ and for a single patche $r_{p_{l}}$, we have
$$\arg\min_{\alpha_l}\big\{{1 \over 2}\|D\alpha_{l}-r_{p_{l}}\|_{\ell_2}^{2}+{\theta_{1,l}}\|\alpha_l\|_{\ell_1}+{\theta_{2,l}}\|\alpha_l-\alpha_{t-l_{l}}^{*}\|_{\ell_1}\big\}, \eqno(33)$$
where $\theta_{\chi}={\rm diag}\big((\theta_{\chi , l})_{l=1}^{J}\big) , \chi=\{1,2\}.$

The sub-problem of (33) can be solved adeptly via an iterative-shrinkage algorithm using surrogate function (see Appendix II, for details). Such this method is based on the work of Daubechies \emph{et al}. [38], and was used proficiently by Dong \emph{et al}. [39] in order to solve a double-header $\ell_{1}$ optimization.

We utilize the solution of (27) into an refinement (post processing) optimization problem in order to enhance the recovery of current frame. Suppose $v^{*}$ is the solution of (27), then the refinement optimization problem can be formulated as:
$$\{u_{t}^{*},\{\alpha_{t_{l}}^{*}\}_{l=1}^{J},D\}=\mathop{\min }_{\alpha_{t}, u_t, D_t}\big\{{1 \over 2}\|f-\Phi u_t\|_{\ell_2}^2+{1 \over 2}\|u_t-v^{*}\|_{\ell_2}^2+\sum_{l=1}^{J}\mu_{l}^{\prime}\|\alpha_{t_{l}}\|_{\ell_0}+{\lambda^{\prime}\over 2}\sum_{l=1}^{J}{\| D_t\alpha_{l}-R_{l}u_{t}}\|_{\ell_q}\big\}, \eqno(34)$$
where $\mu_{l}>0$ are some regularization parameters that control the image patch sparsity. Besides, $\lambda^{\prime}$ is weight parameter controls the trade-off between the data fidelity and the image prior. Here, $D_{t}$ is the ALS basis of current frame that can be learned (initially using $D$ and $v^{*}$) and updated via the methods of [30]-[32]. Indeed, the second term measures the distance between the estimated frame and the enhanced one. The problem of (34) can be solved in an iterative mode, by decoupling it into three sub-problems of \emph{sparse coding}, \emph{dictionary learning}, and \emph{reconstruction}. The sparse coding problem can be solved using OMP, and $D_t$, in dictionary learning step, updates by the methods of [30]-[32]. Also, the reconstruction step has a closed-form solution
$$u_{t}^{k+1}=(\Phi\Phi^{T}+I+\lambda^{\prime}\sum_{l=1}^{J}R_{l}^{T}R_{l})^{-1}(v^{*}+\Phi^{T}f+\lambda^{\prime}\sum_{l=1}^{J}R_{l}^{T}D\alpha_{l}). \eqno(35)$$

At a glance, it seems that (34) is very similar to that proposed in [40], however, in fact we use (34) in order to update the variables which used for the recovery of next frame, additionally with a minor enhancement on the recovered frame using (27). 
\section{Eperimental Results}
In this section, we evaluate the performance of the proposed method.
To evaluate our simulation results, we use two applicable quality assessors, the peak signal-to-noise (PSNR) in dB and the structural similarity (SSIM) [41]. For an 8 bits gray scale $I_{r}\times I_{c}$ image (frame), PSNR is calculated as$$PSNR=20\log_{10}{\sqrt{I_{r}\times I_{c}} 255\over \|u-\tilde{u}\|}, \eqno(36)$$where $u$ and $\tilde{u}$ denote the original image and the reconstructed image, respectively.

SSIM [41] is another assessor for measuring the similarity between two images, and it is near to human eye perception. Here, the obtained results are evaluated by a specific form of the SSIM index:$$SSIM(u,\tilde{u})={(2\mu_{u} \mu_{\tilde{u}} +c_{1})(2{\sigma}_{u{\tilde{u}}}+c_{2})\over ({\mu_u}^{2}+{\mu}_{\tilde{u}}^2+c_{1})({\sigma}_{u}^{2}+{\sigma}_{\tilde{u}}^{2}+c_{2})} \eqno(37)$$where $\mu_{u}$ ($\mu_{\tilde u}$) and $\sigma_{u}$ ($\sigma_{\tilde u}$) denote the average and variance of $u$ ($\tilde u$), respectively. $\sigma_{u\tilde{u}}$ is the covariance of $u$ and $\tilde{u}$. $c_{1}$ and $c_{2}$ are two variables to stabilize the division with weak denominator, such that $c_{1}=(k_{1} L)^{2}, c_{2}=(k_{2} L)^{2}$ and $0<k_{1},k_{2}\ll1$. $L$ is the dynamic range of pixel values (255 for 8 bits gray scale images). For our experiments, we use $k_{1}=0.01$ and $k_{2}=0.03$. All experiments were performed using MATLAB 2013a, on a computer equipped with Intel$\circledR$ core$^{TM}$ i7, 3.7 GHz processor, with 48 GB of RAM, and running on Windows 7. The performances of our experiments are evaluated on the luminance component of four well-known video test sequences (e.g., ``\emph{Foreman}", ``\emph{Coastguard}", ``\emph{Mobile and Calendar}" and ``\emph{Hall Monitor}") with a CIF resolution of $352\times288$ pixels. Also, in all experiments we use the block size of $32 \times 32$ and the size of each patch is set to $8 \times 8$. The CS measurement of each blocks is obtained by applying an orthonormalized i.i.d. Gaussian projection matrix to each of them, however the orthonormality of the projection matrix makes the solution of problems more simpler. Note that, in this paper we do not consider the quantization and entropy encoding of measurements, since they are beyond the scope of this paper.
\subsection{Experiment 1}
\begin{figure}
         {\centering{
         \begin{subfigure}[b]{0.35\textwidth}
                 \includegraphics[scale=0.3]{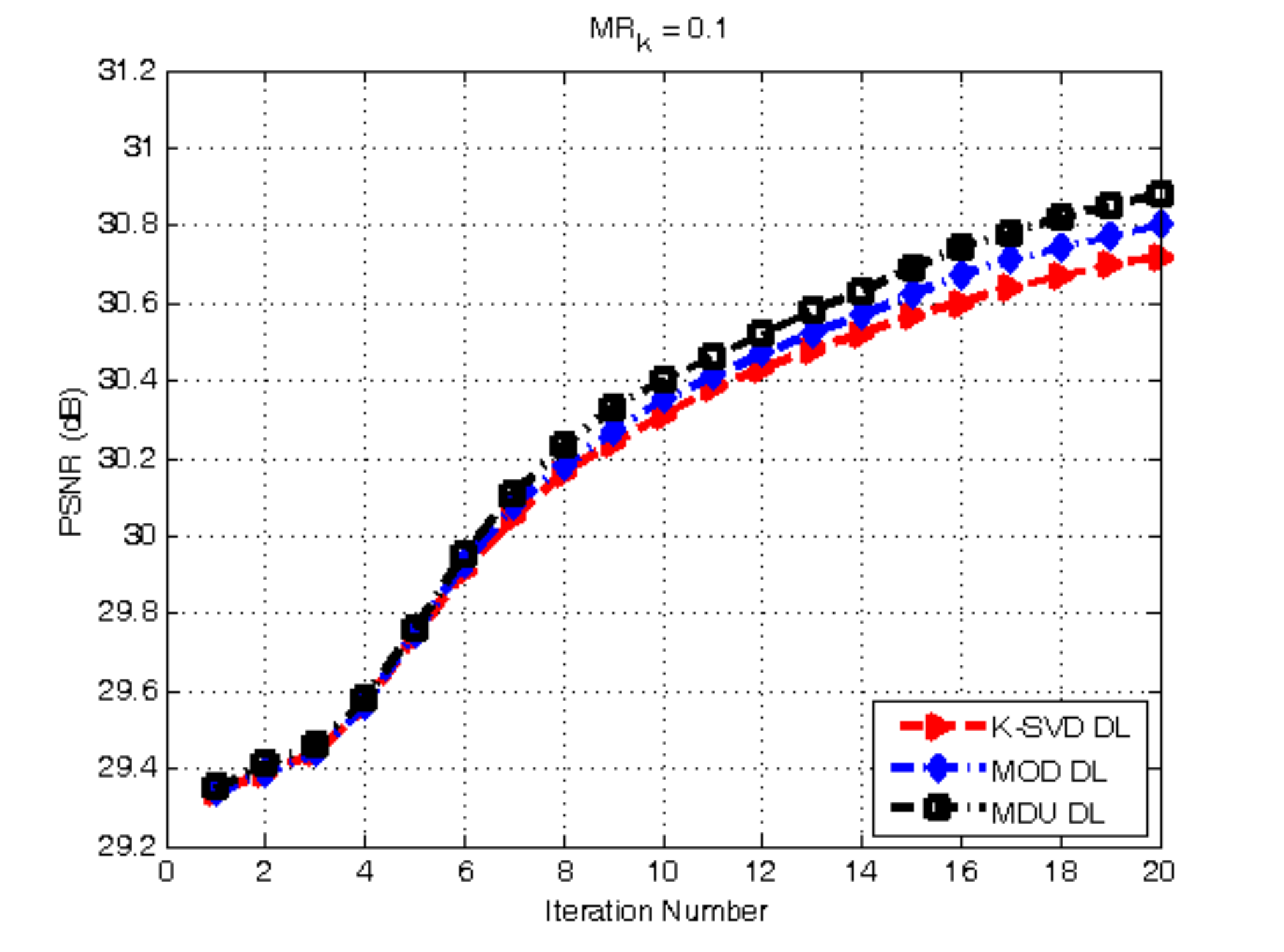}
                  \centering
                 \caption{}
                 \label{fig:1}
         \end{subfigure}%
         \begin{subfigure}[b]{0.33\textwidth}
                 \includegraphics[scale=0.3]{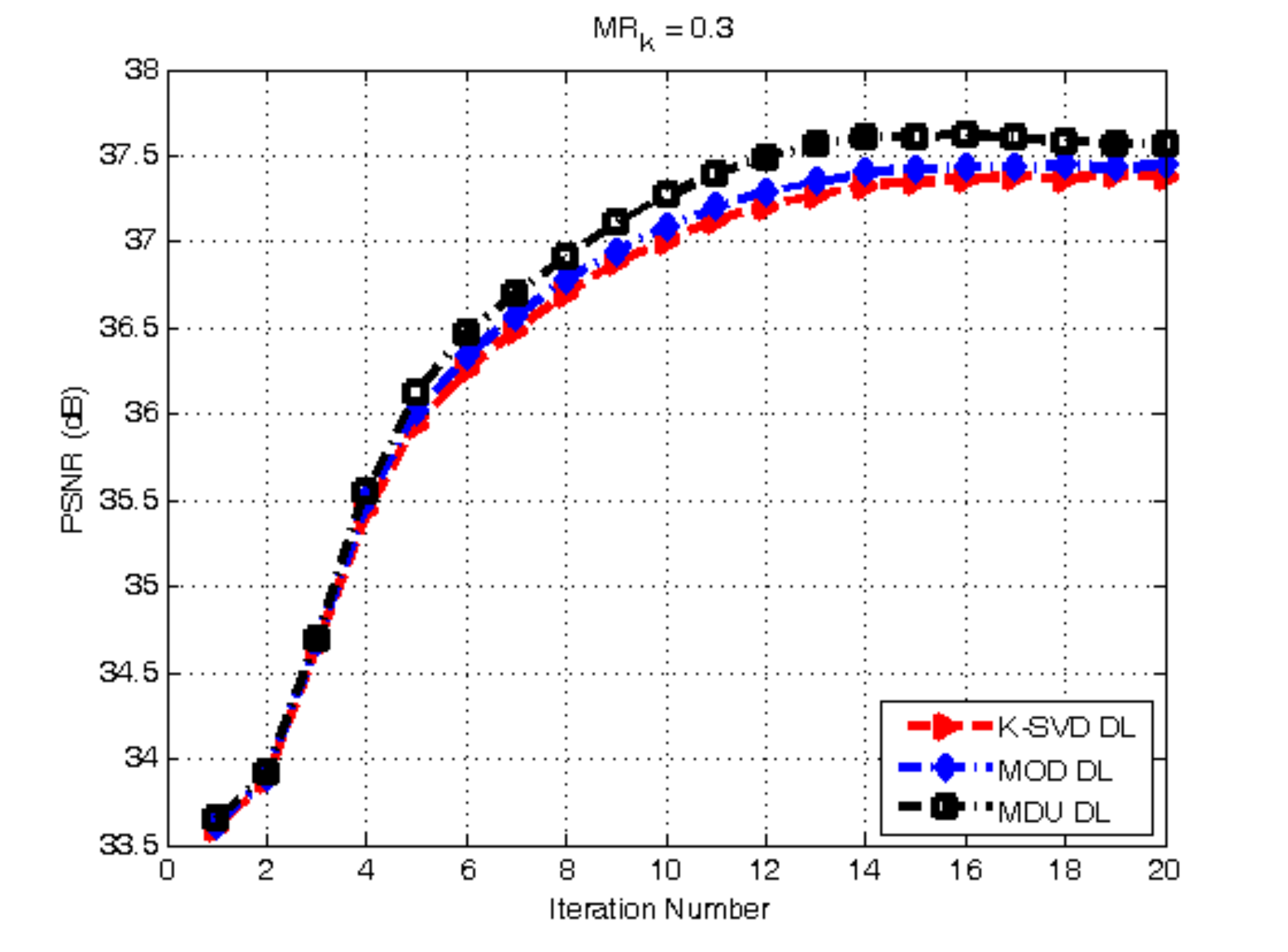}
                 \centering
                 \caption{}
                 \label{fig:1}
         \end{subfigure}
         \begin{subfigure}[b]{0.3\textwidth}
                 \includegraphics[scale=0.3]{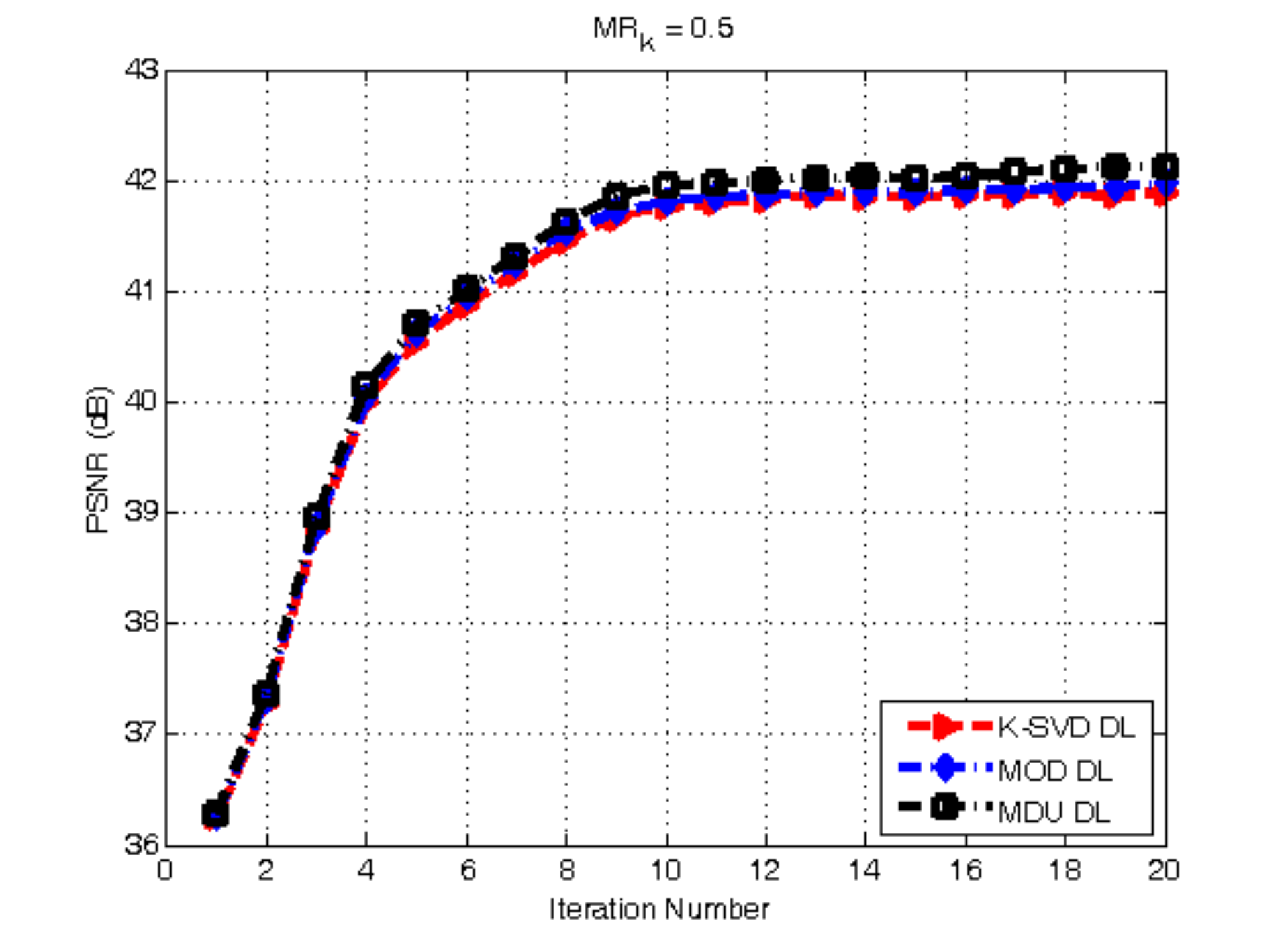}
                 \centering
                 \caption{}
                 \label{fig:1}
         \end{subfigure}\\}}
         {\footnotesize{Fig. 1: Performance of different dictionary learning algorithms for the proposed method, on the frame 21\emph{st} of \emph{Foreman} sequence with various $MR_{K}$. (a) $MR_{K}=0.1$: ${\bar{t}}_{K-SVD}=11.41$ sec, ${\bar{t}}_{MOD}=12.57$ sec, ${\bar{t}}_{MDU}=112.33$ sec. (b) $MR_{K}=0.3$: ${\bar{t}}_{K-SVD}=15.1$ sec, ${\bar{t}}_{MOD}=16.05$ sec, ${\bar{t}}_{MDU}=169.8$ sec. (c) $MR_{K}=0.5$: ${\bar{t}}_{K-SVD}=16.96$ sec, ${\bar{t}}_{MOD}=17.78$ sec, ${\bar{t}}_{MDU}=203.25$ sec.}}
\end{figure}

In the first experiment, we evaluate the effectiveness of the discussed dictionary learning methods of [30]-[32] for the proposed key frame recovery algorithm. Though, the other dictionary learning methods can be investigated to adopt in our scheme. The key frame is reconstructed initially using the method of [18], in order to use as the initial training image in the process of learning an ALS basis (dictionary). The ALS basis is obtained by each of the methods provided in [30]-[32], in which the default parameter setting is as follows: the size of sparsifying basis (dictionary) is 256 and number of training iteration is 20. Also, in corresponding recovery problems (see Table II), $iin=200$, $k_{max}=20$, $Tol=10^{-4}$, $\mu$ is set to be $2.5\times 10^{-3}$ and $\lambda$ is set empirically. In Fig. 1, a twenty-time-iteration of our method is illustrated as an example to show the performance of MOD [30], K-SVD [31] and MDU [32] as the dictionary learning methods, for recovery of the 21\emph{st} frame of the \emph{Foreman} with different $MR_{K}$. 
In Fig. 1, $\bar{t}$ shows the average time of dictionary learning process at each iteration. Obviously, the results show that MDU provides a better recovery performance (in quality) compared to the other mentioned methods, but with the cost of higher computational complexity. The extra cost is only generated from the dictionary update step in each iteration. Also, the PSNR performance of K-SVD and MOD are very close to each other, but the first one gains lower computational complexity. So, it seems reasonable to adopt K-SVD as the dictionary learning algorithm in our method. The same experiments were run on \emph{Coastguard}, \emph{Hall Monitor} and \emph{Mobile and Calendar} sequences (with fast motion scene); the obtained results proved the accuracy of our assumption.
\begin{figure}
         {\centering{
         \begin{subfigure}[b]{0.3\textwidth}
                 \includegraphics[scale=0.32]{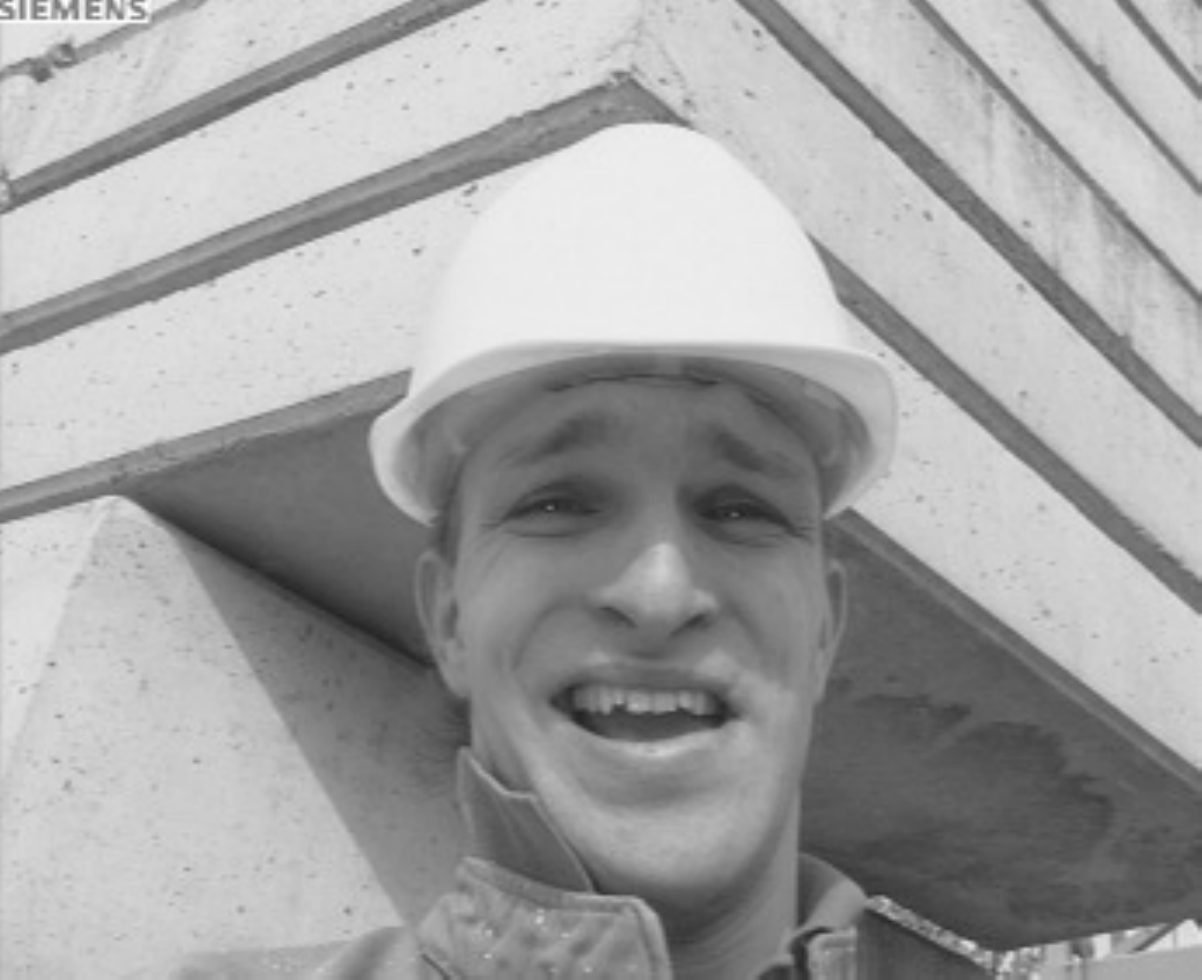}
                 \centering
                 \caption{}
                 \label{fig:1}
         \end{subfigure}
         \begin{subfigure}[b]{0.3\textwidth}
                 \includegraphics[scale=0.32]{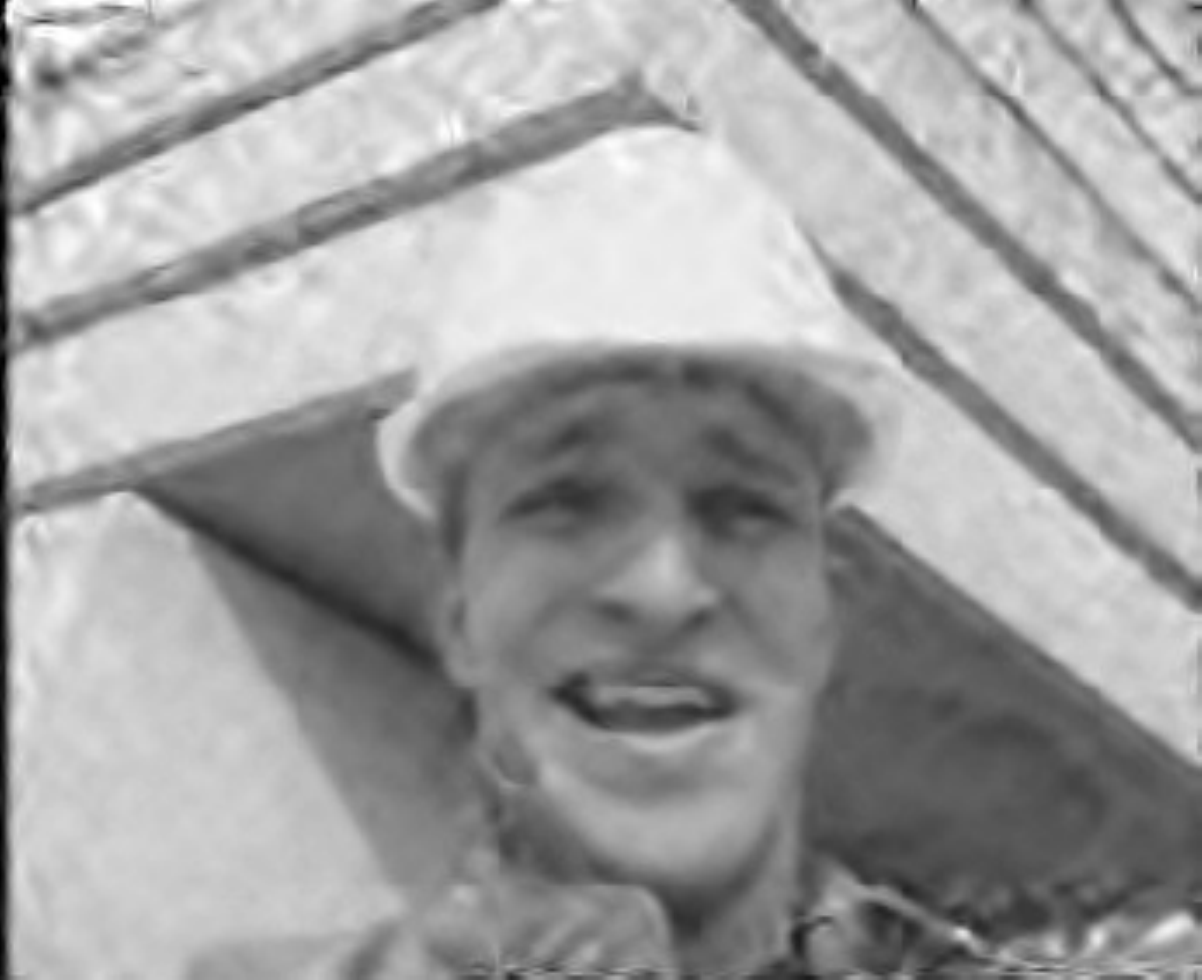}
                 \centering
                 \caption{}
                 \label{fig:1}
         \end{subfigure}%
         \begin{subfigure}[b]{0.316\textwidth}
                 \includegraphics[scale=0.32]{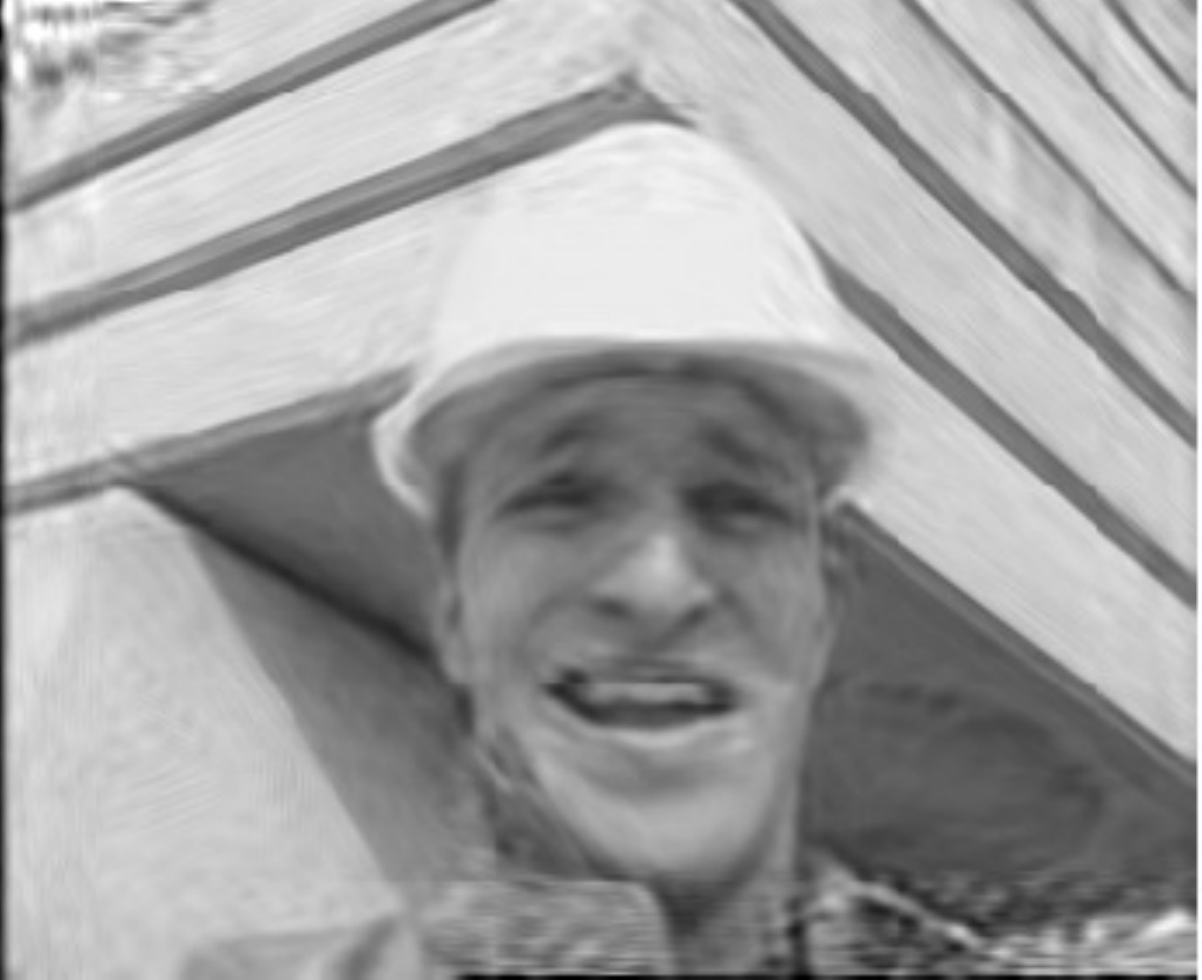}
                 \centering
                 \caption{}
                 \label{fig:1}
         \end{subfigure}}}{\centering{
         \begin{subfigure}[b]{0.3\textwidth}
                 \includegraphics[scale=0.32]{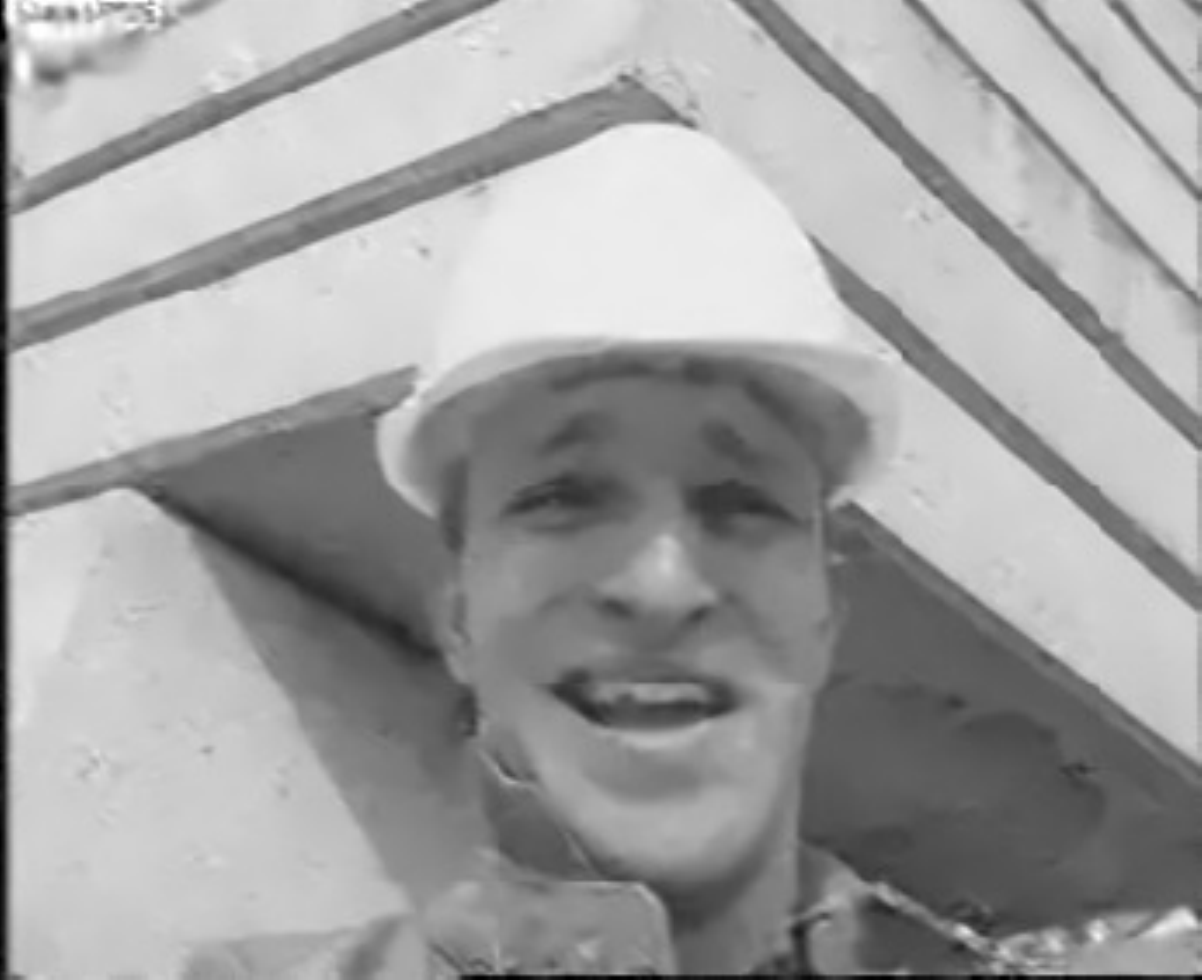}
                 \centering
                 \caption{}
                 \label{fig:1}
         \end{subfigure}
         \begin{subfigure}[b]{0.295\textwidth}
                 \includegraphics[scale=0.32]{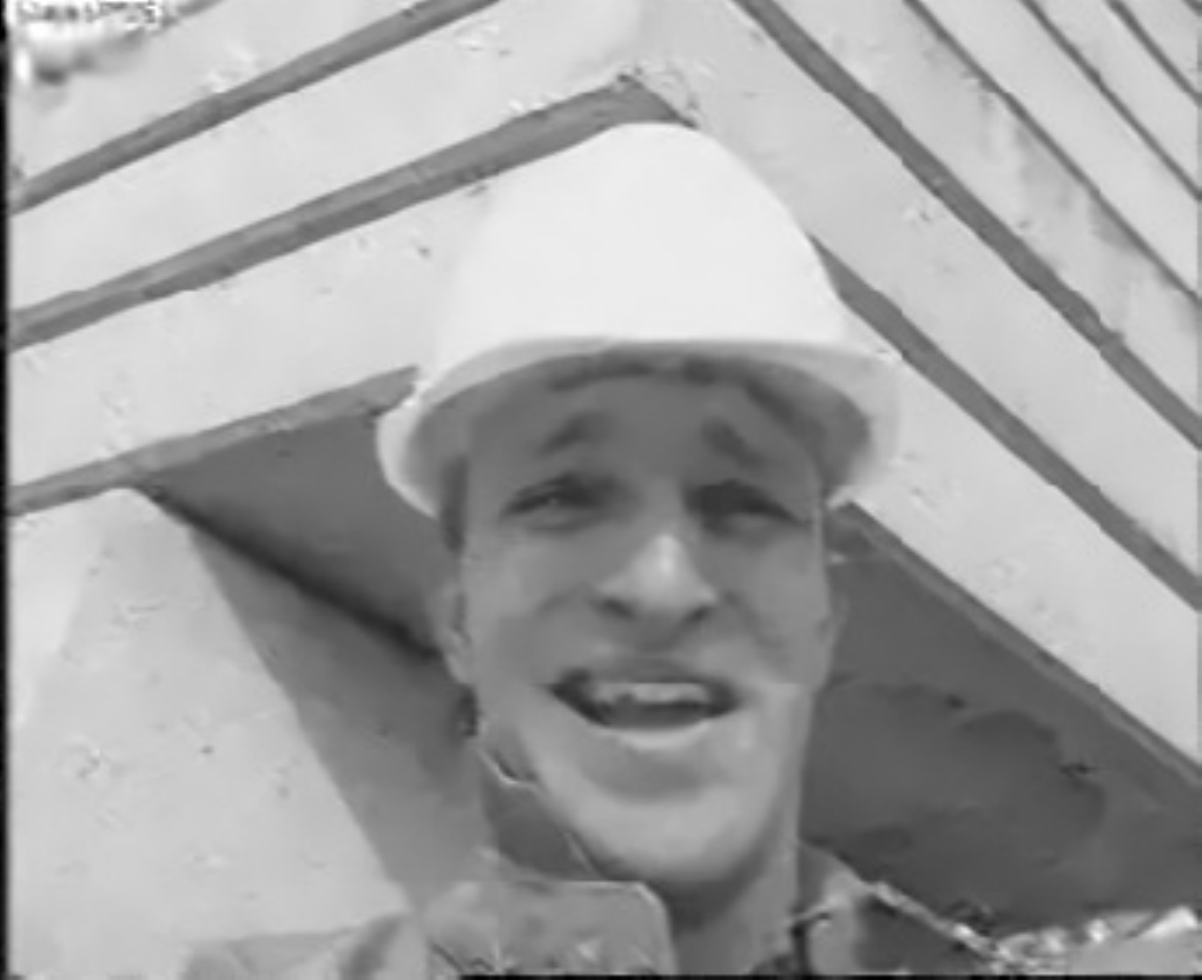}
                 \centering
                 \caption{}
                 \label{fig:1}
         \end{subfigure}
         \begin{subfigure}[b]{0.31\textwidth}
                 \includegraphics[scale=0.32]{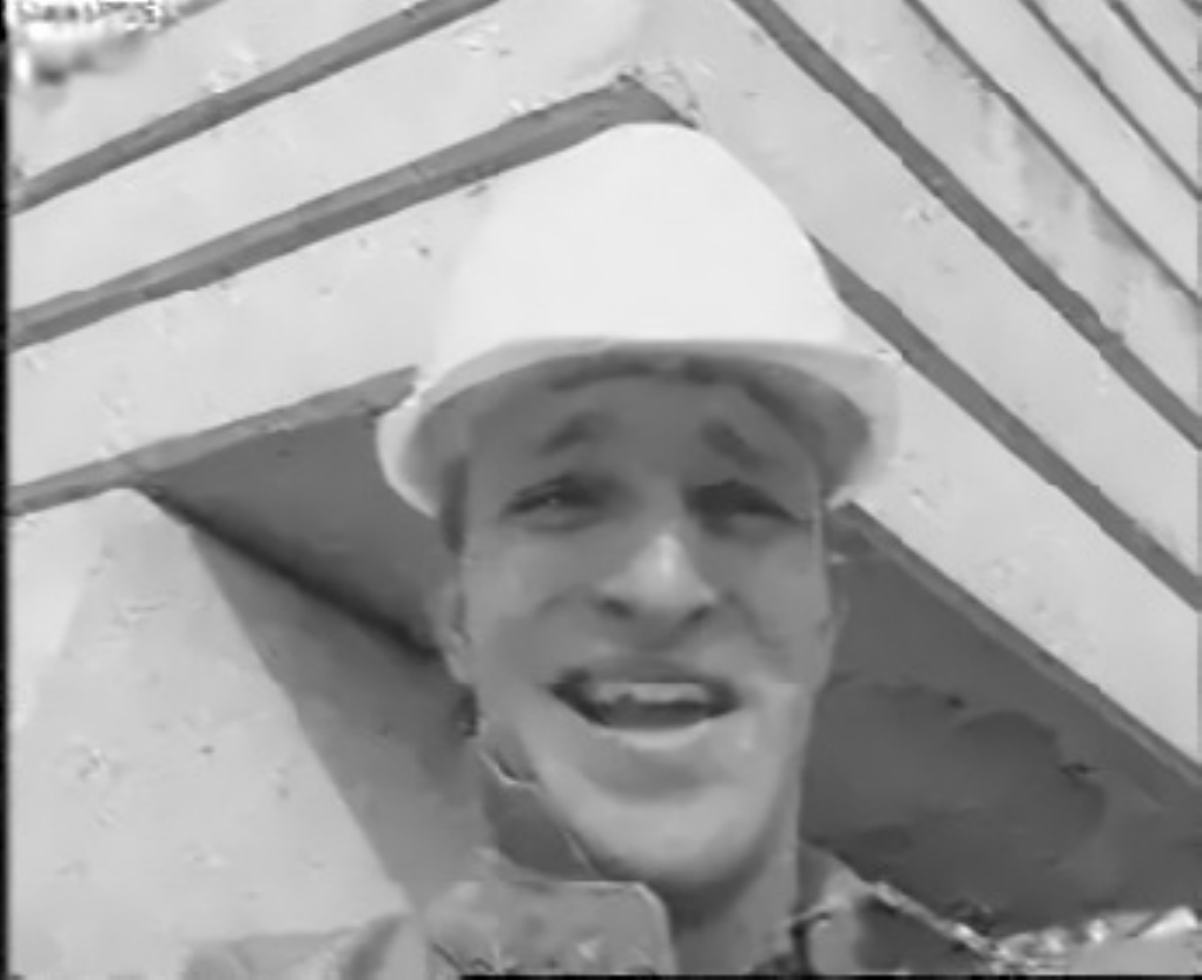}
                 \centering
                 \caption{}
                 \label{fig:1}
         \end{subfigure}\\}}
         {\footnotesize{Fig. 2: Different decodings of the 21\emph{st} frame of \emph{Foreman} with respect to recovery as a key frame ($MR_{K}=0.1$). (a) Original frame. Reconstructed 21\emph{st} frame by usnig (b) The 2D DDWT basis (PSNR=26.82 dB, SSIM=0.783). (c) The multi-hypothesie method [11] (PSNR=28.93 dB, SSIM=0.824). The proposed method via: (d) K-SVD (PSNR=30.72 dB, SSIM=0.875). (e) MOD (PSNR=30.8 dB, SSIM=0.876). (f) MDU (PSNR=30.88 dB, SSIM=0.878).}}
\end{figure}
Fig. 2 shows the different decoding of the 21\emph{st} frame of the \emph{Foreman}, produced by 2D DDWT [42] basis intra-frame decoder (with 3 level of decomposition) [Fig. 2(b)], the MH [18] decoder [Fig. 2(c)] and the proposed key frame recovery method using K-SVD [Fig. 2(d)], MOD [Fig. 2(e)] and MDU [Fig. 2(f)]. Note that, for fair comparison, the same test conditions (i.e., the same sensing matrix) are used in all experiments and all experimental results are averaged over 5 independent trials. Evidently, it can be observed that the fixed basis intra-frame decoder and MH-based decoder suffer noticeable performance loss over the whole image, while the proposed key frame recovery decoder demonstrates a considerable reconstruction quality improvement. 

\subsection{Experiment 2}
In this experiment, we evaluate the performance of the proposed method (via K-SVD dictionary learning) for decoding of the first 50 frames of the test video sequences, with different measurement ratio $MR$ scenario. More specifically, here, two proposed CVS decoders are examined for all four test sequences, i.e., when $MR_{k}=MR_{NK}$, and when $MR_{NK}\leq MR_{K}$= 0.5. Also, for comparision purposes, two existing typical CVS decoders is considered, e.g., the 2D DDWT basis intra-frame decoder, and the MH predictions inter-frame decoder [18]. The group of the picture (GOP) size is set to 5 and the default parameter setting is as well as experiment 1. The number of iterations in algorithm, $k_{max}$, is set to 6; it is clear that the higher number of iteration yields slightly better performance in quality of recovered sequence, but at the cost of higher execution times of the algorithm. It is worth emphasizing that, however by increasing the number of iterations, the quality of recovered frame increases, but after some iterations it yields an insignificant improvement, i.e., see Fig. 1(c) that PSNR values after the 10\emph{th} iteration converge.
\begin{figure}
         {\centering{
         \begin{subfigure}[b]{0.35\textwidth}
                 \includegraphics[scale=0.42]{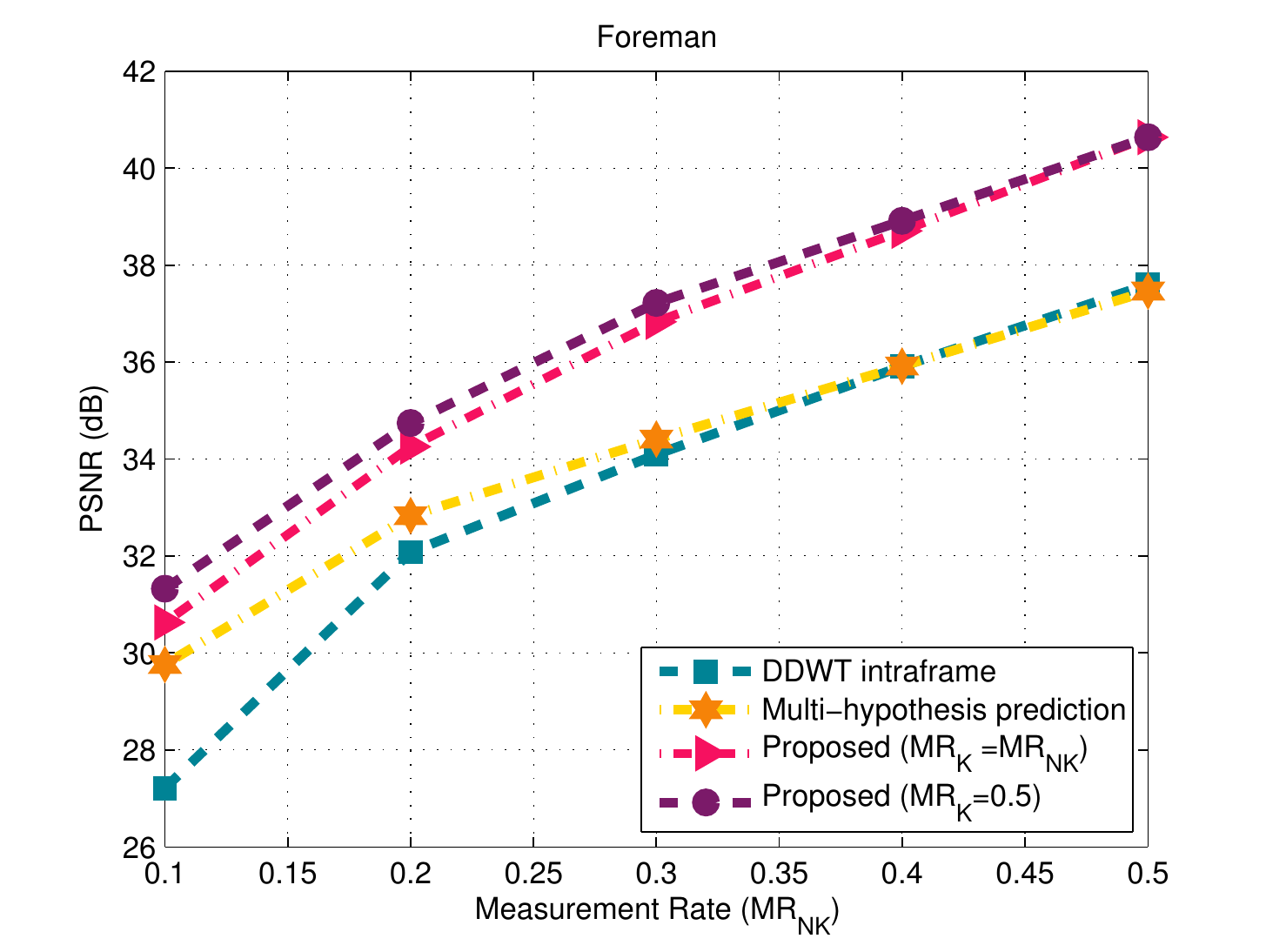}
                  \centering
                 \caption{}
                 \label{fig:1}
         \end{subfigure}%
         \begin{subfigure}[b]{0.35\textwidth}
                 \includegraphics[scale=0.42]{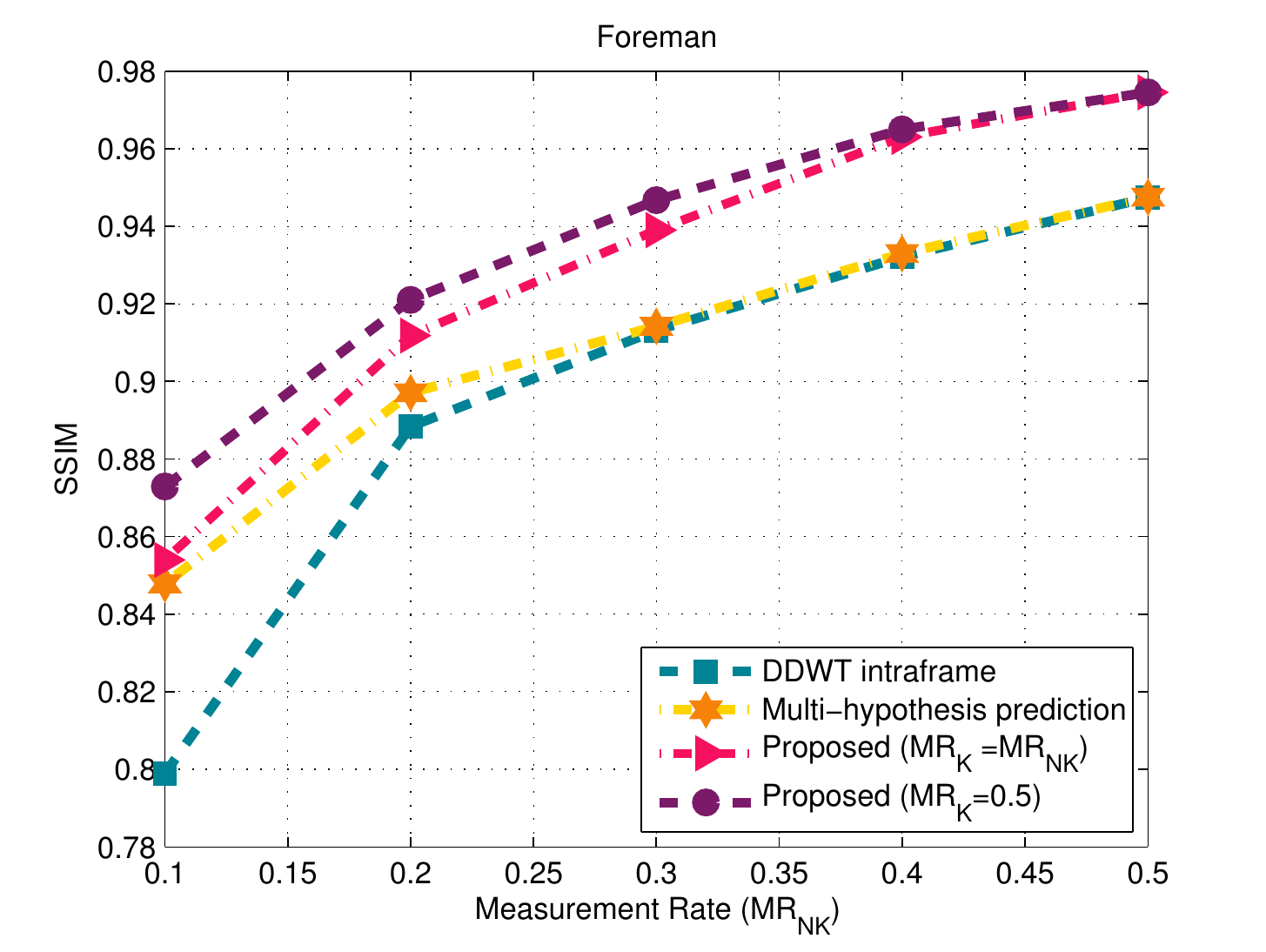}
                 \centering
                 \caption{}
                 \label{fig:1}
         \end{subfigure}\\
         \begin{subfigure}[b]{0.35\textwidth}
                 \includegraphics[scale=0.42]{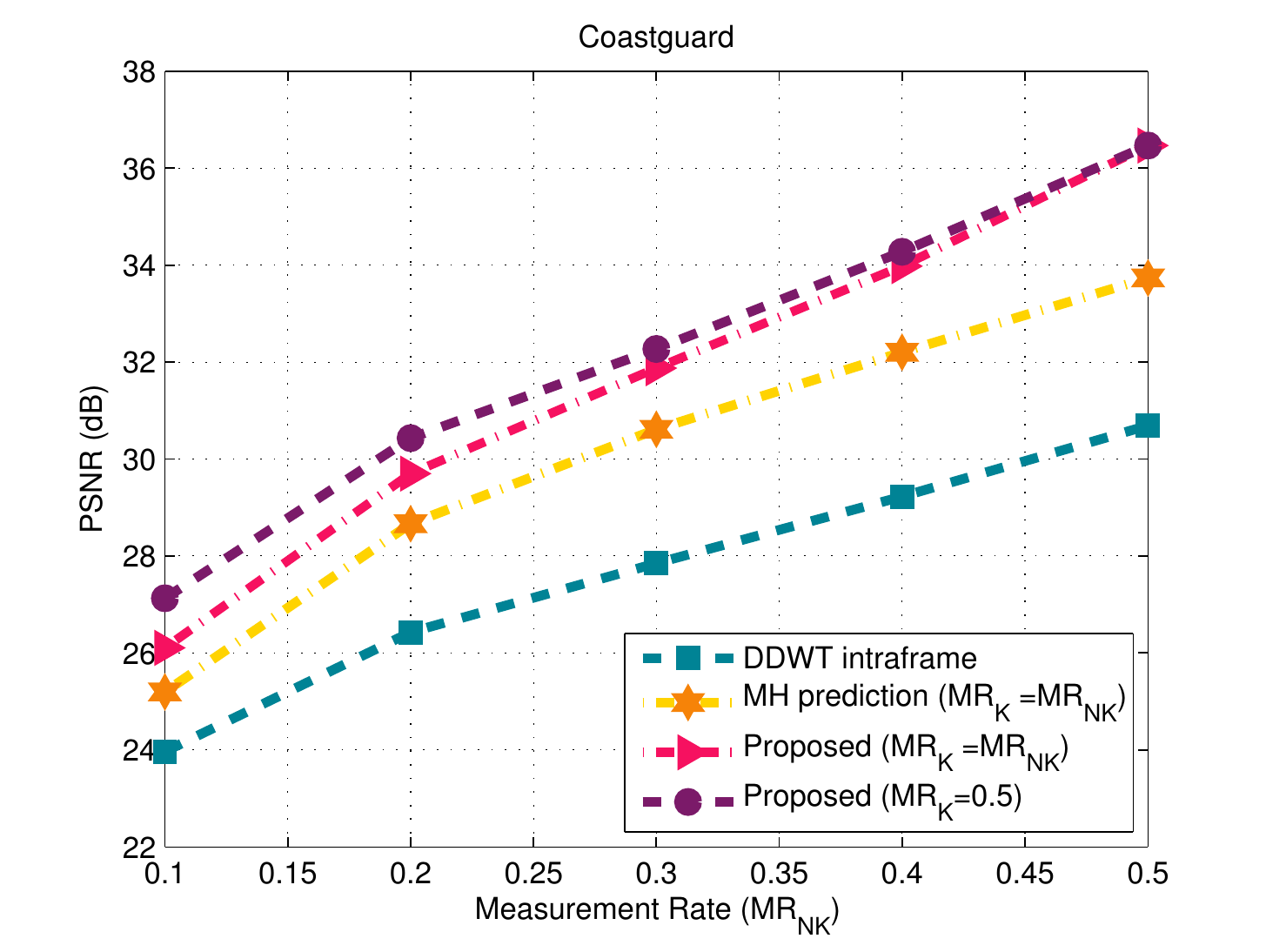}
                  \centering
                 \caption{}
                 \label{fig:1}
         \end{subfigure}%
         \begin{subfigure}[b]{0.35\textwidth}
                 \includegraphics[scale=0.42]{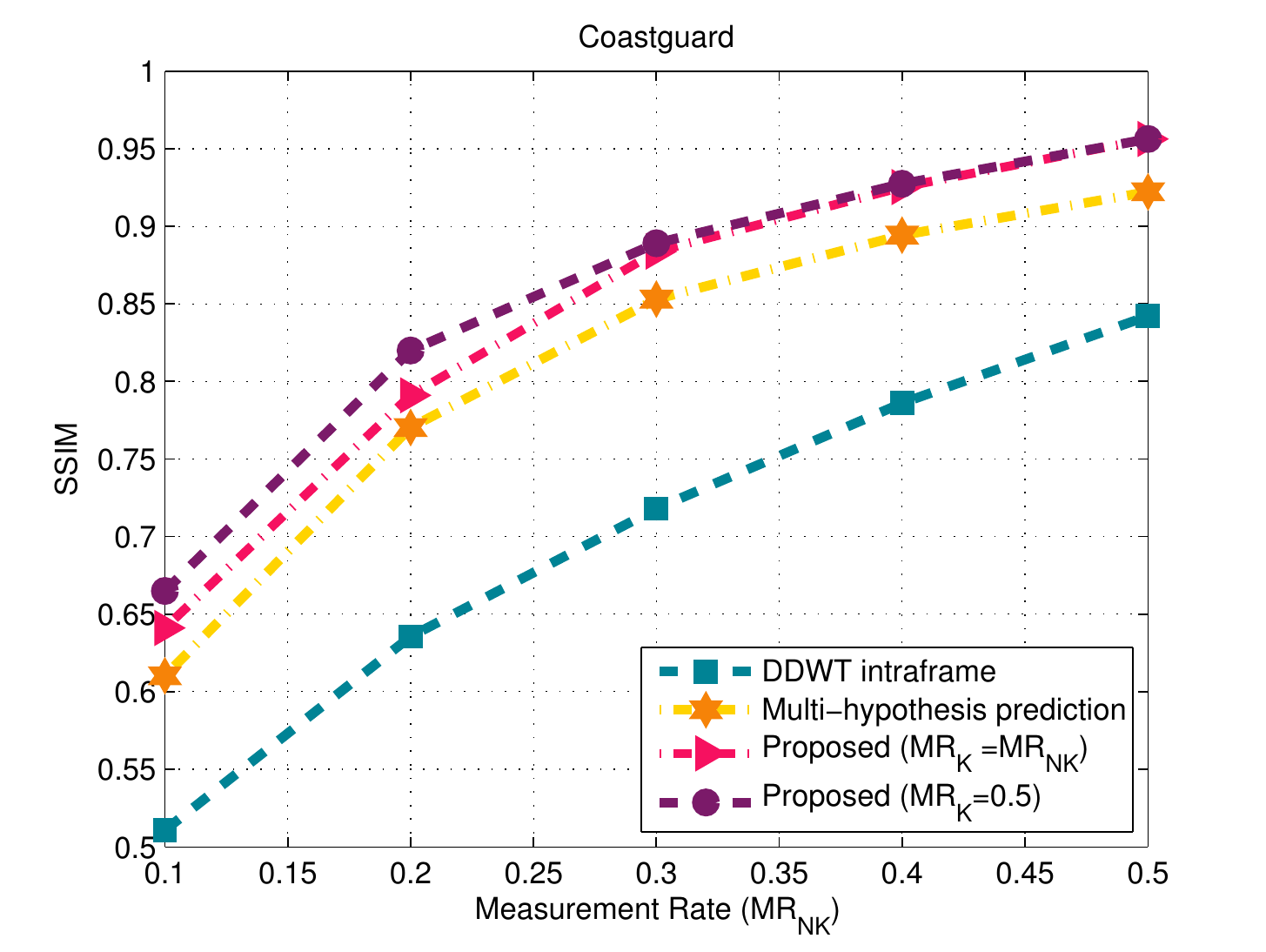}
                 \centering
                 \caption{}
                 \label{fig:1}
         \end{subfigure}\\}}
         {\footnotesize{Fig. 3: The PSNR and SSIM performance of proposed method on the first 50 frames of the \emph{Foreman} and the \emph{Coastguard} sequence. \emph{Foreman}: (a) PSNR vs. MR. (b) SSIM vs. MR. \emph{Coastguard}: (c) PSNR vs. MR. (d) SSIM vs. MR.}}
\end{figure}

In this experimental studies, Fig. 3 shows the PSNR and SSIM performance of the proposed method in various measurement ratios, compared with the other mentioned recovery algorithms for the \emph{Foreman} and the \emph{Coastguard} video sequences. The PSNR and SSIM values shown in Fig. 3, are averaged over all PSNR and SSIM values of the reconstructed frames.
Fig. 3(a) and Fig. 3(b) show the PSNR and SSIM performance of the proposed method, respectively, compared with the 2D DDWT intra-frame and the MH inter-frame recovery method ($MR_{K}=MR_{NK}$), in various measurement ratios. Figs. 3(a)-(b) are obtained for the \emph{Foreman} sequence, while Figs. 3(c)-(d) show the same scenario as used in Figs. 3(a)-(b), but for the \emph{Coastguard} sequence. As can be seen, the proposed method outperforms significantly in both PSNR and SSIM values, compared to the fixed basis intra-frame and MH inter-frame decoders. Figs. 4 and 5 show the decoding comparisions with different measurement ratios scenario for the 49\emph{th} frame of the \emph{Coastguard}, 40\emph{th} frame of the \emph{Hall Monitor}, and 30\emph{th} frame of the \emph{Mobile and Calendar}, respectively. The frames used in Figs. 4 and 5 are all regarded as the non-key frames. Figs. 4 and 5 (d)-(f) depict the magnitude of error reconstruction (error map). Differences among the error maps are clearer with zoom-in. The error maps show the lower errors obtained by the proposed method compared to the other mentioned methods. Also, the averaged PSNR and SSIM values of all first 50 frames of the \emph{Mobile and Calendar} and the \emph{Hall Monitor} are listed in details in Table III.

Obviously, the numerical results show that, the proposed method (in both states, $MR_{K}$ = $MR_{NK}$ and $MR_{NK}\leq MR_{K}$ = 0.5) gains better performance, in both PSNR and SSIM, compared with the other mentioned methods. Also note that, by increasing $MR_{K}$, the reconstruction quality will increase, but poor compression is its consequence.  

\begin{figure}
         {\centering{
         \begin{subfigure}[b]{0.191\textwidth}
                 \includegraphics[scale=0.265]{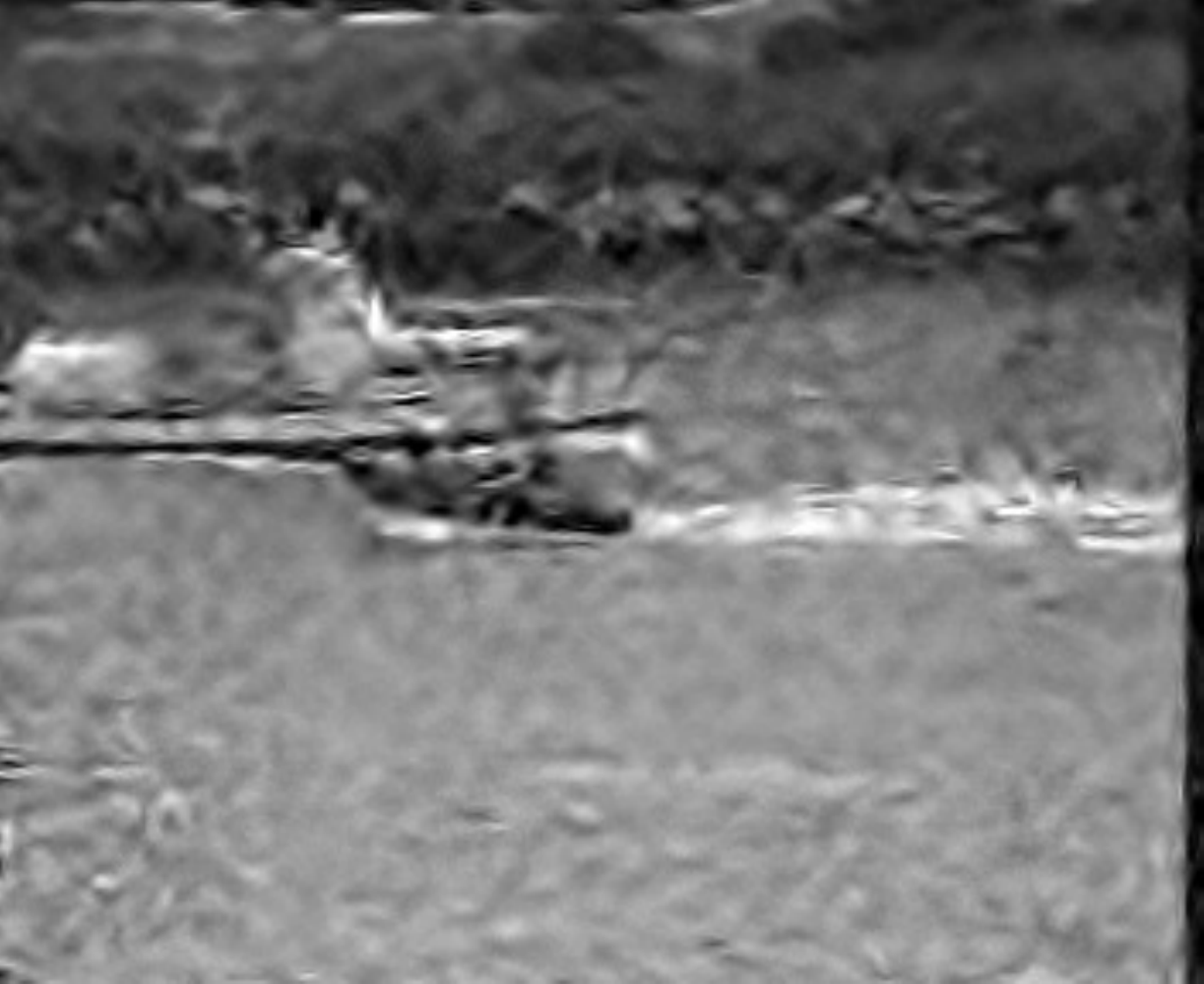}
                 \centering
                 \caption{}
                 \label{fig:1}
         \end{subfigure}
         \begin{subfigure}[b]{0.21\textwidth}
                 \includegraphics[scale=0.265]{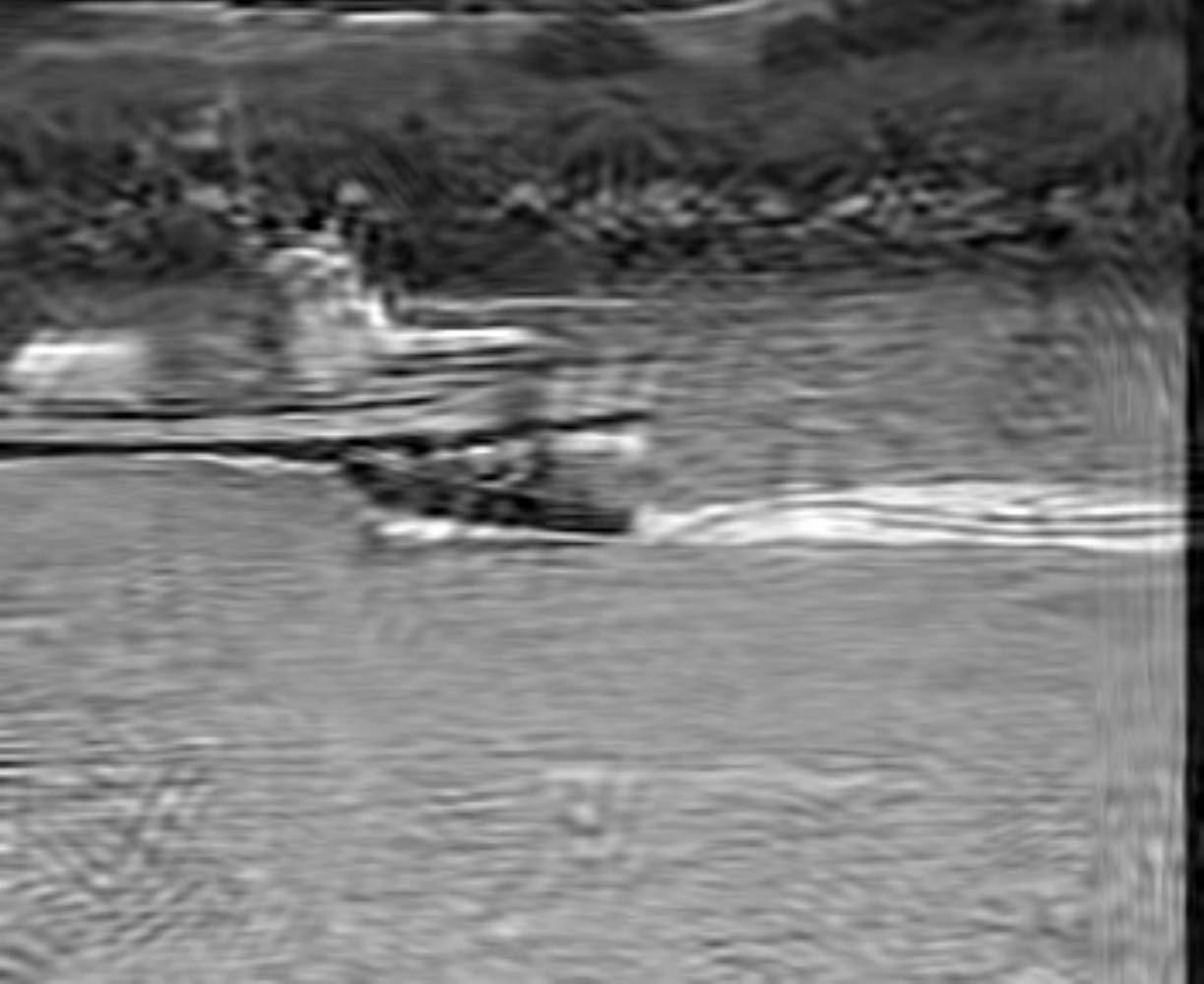}
                 \centering
                 \caption{}
                 \label{fig:1}
         \end{subfigure}%
         \begin{subfigure}[b]{0.19\textwidth}
                 \includegraphics[scale=0.265]{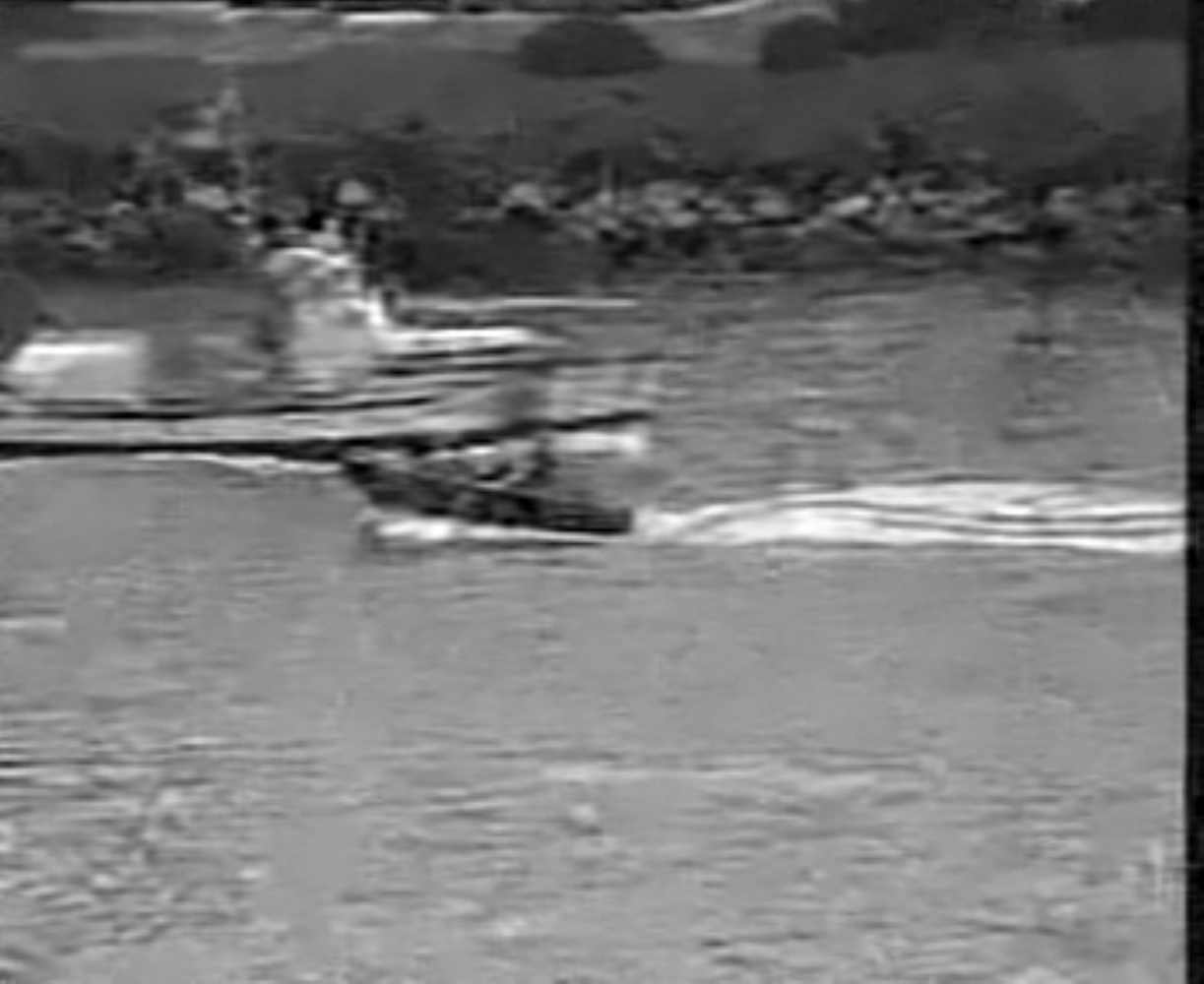}
                 \centering
                 \caption{}
                 \label{fig:1}
         \end{subfigure}\\
         \begin{subfigure}[b]{0.2\textwidth}
                 \includegraphics[scale=0.33]{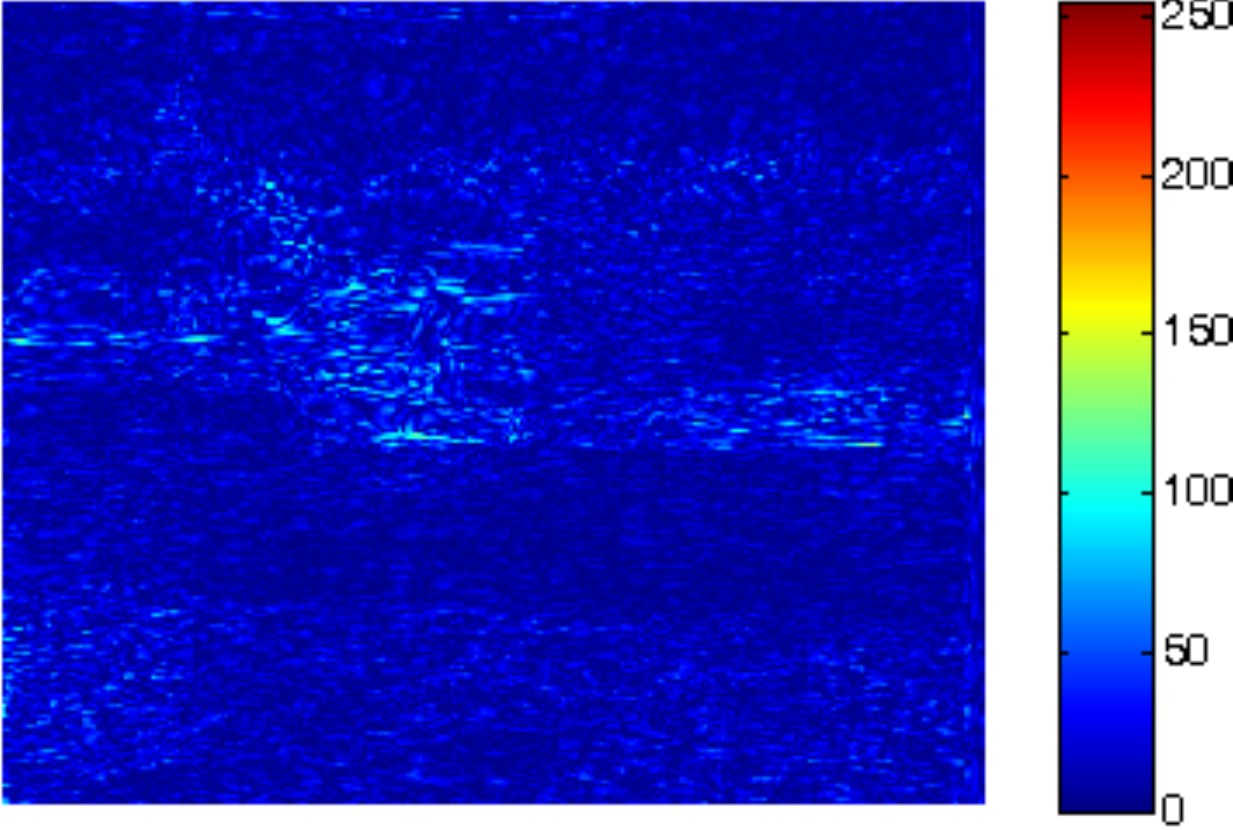}
                 \centering
                 \caption{}
                 \label{fig:1}
         \end{subfigure}
         \begin{subfigure}[b]{0.21\textwidth}
                 \includegraphics[scale=0.33]{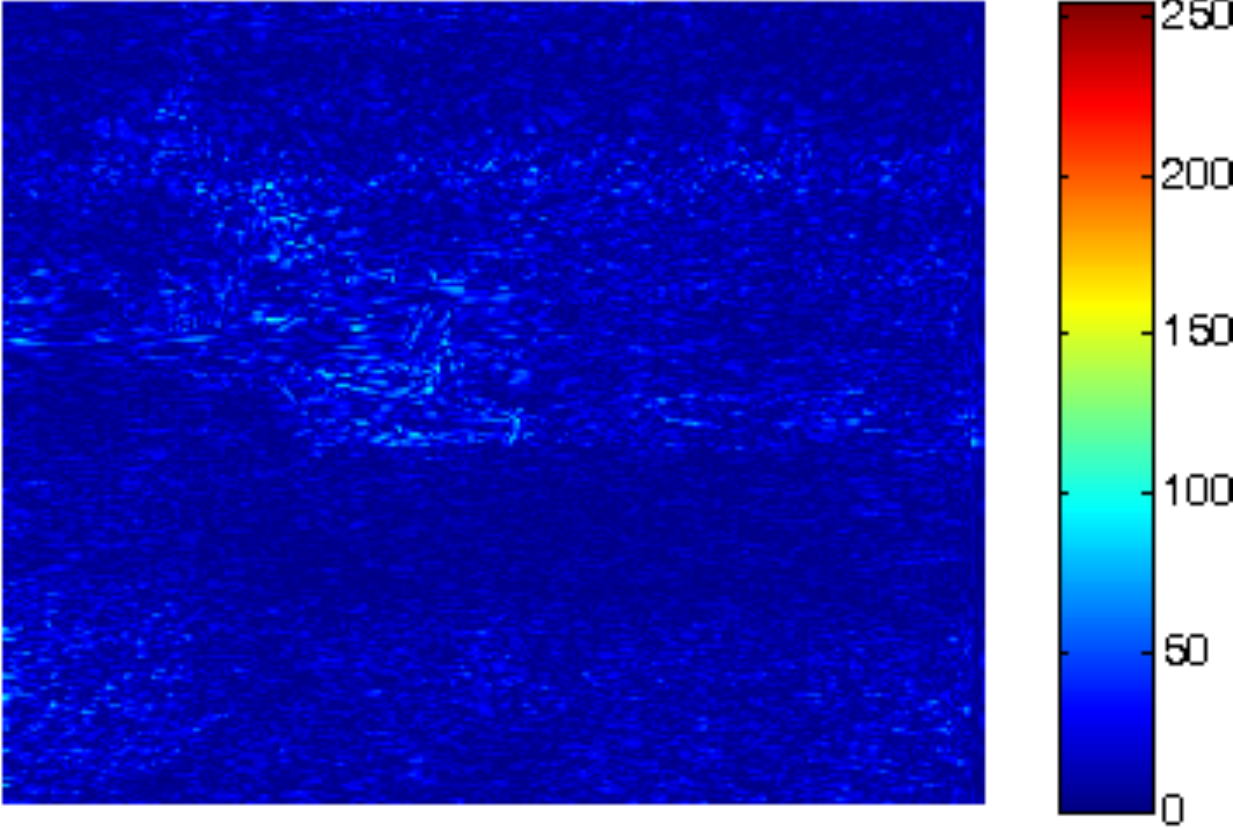}
                 \centering
                 \caption{}
                 \label{fig:1}
         \end{subfigure}%
         \begin{subfigure}[b]{0.2\textwidth}
                 \includegraphics[scale=0.33]{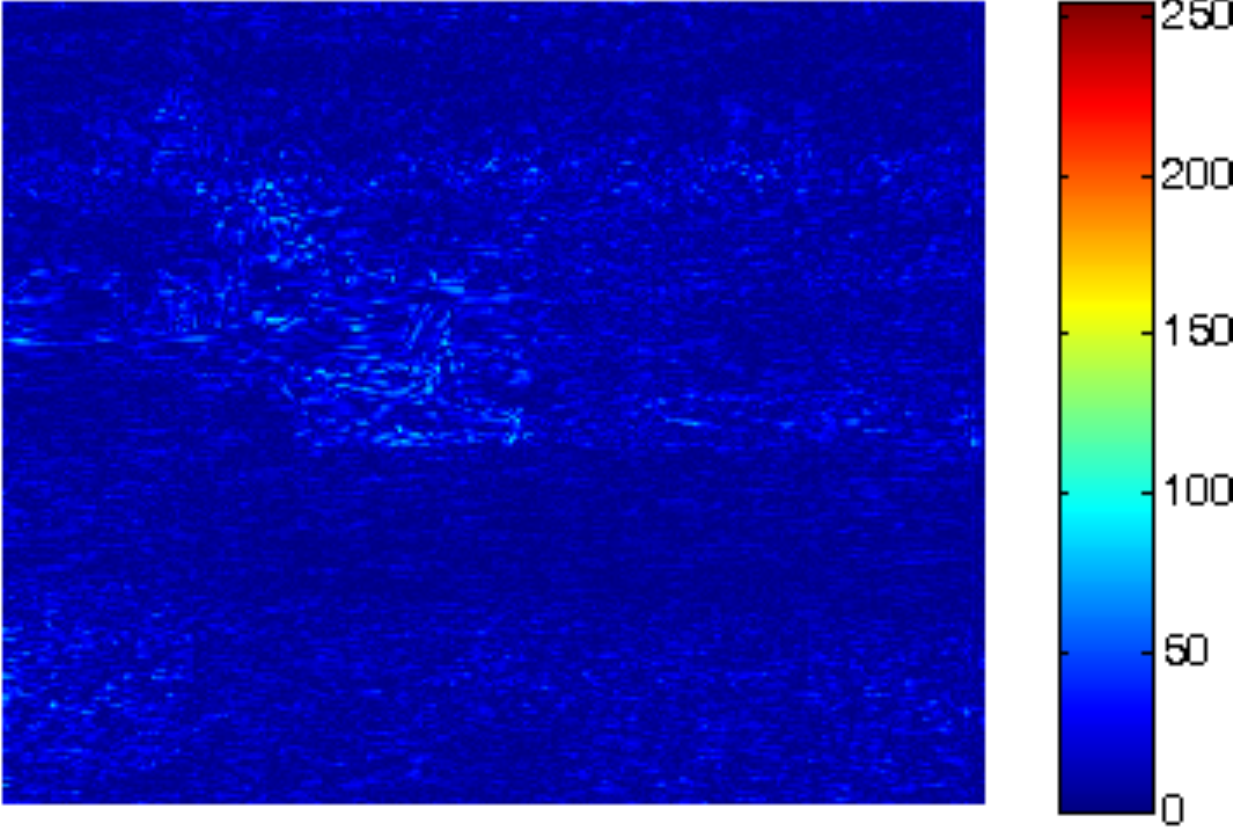}
                 \centering
                 \caption{}
                 \label{fig:1}
         \end{subfigure}\\}}
         {\footnotesize{Fig. 4: Different decoding of the 49\emph{th} frame of \emph{Coastguard} (w.r.t. recovery as a non-key frame $MR_{K}=0.5,MR_{NK}=0.3$). (a) Using the 2D DDWT basis intra-frame decoder (PSNR=23.02 dB, SSIM=0.518). (b) Using the MH method [18] (PSNR=24.38 dB, SSIM=0.630). (c) Using the proposed method (PSNR=25.30 dB, SSIM=0.659). (d)-(f) Magnitude of error reconstruction for (a)-(c), respectively.}}
\end{figure}

\begin{figure}
         {\centering{
         \begin{subfigure}[b]{0.191\textwidth}
                 \includegraphics[scale=0.265]{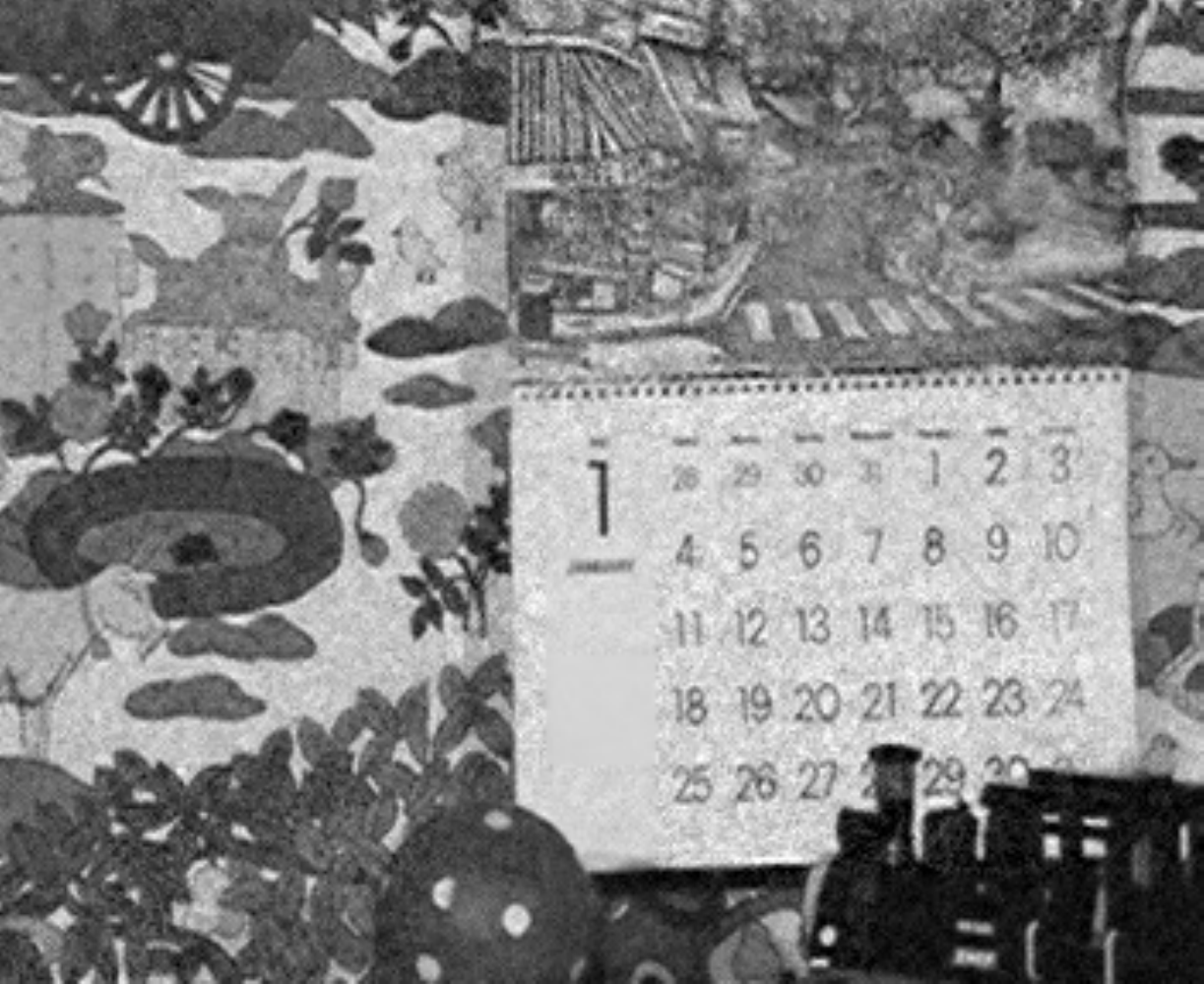}
                 \centering
                 \caption{}
                 \label{fig:1}
         \end{subfigure}
         \begin{subfigure}[b]{0.21\textwidth}
                 \includegraphics[scale=0.265]{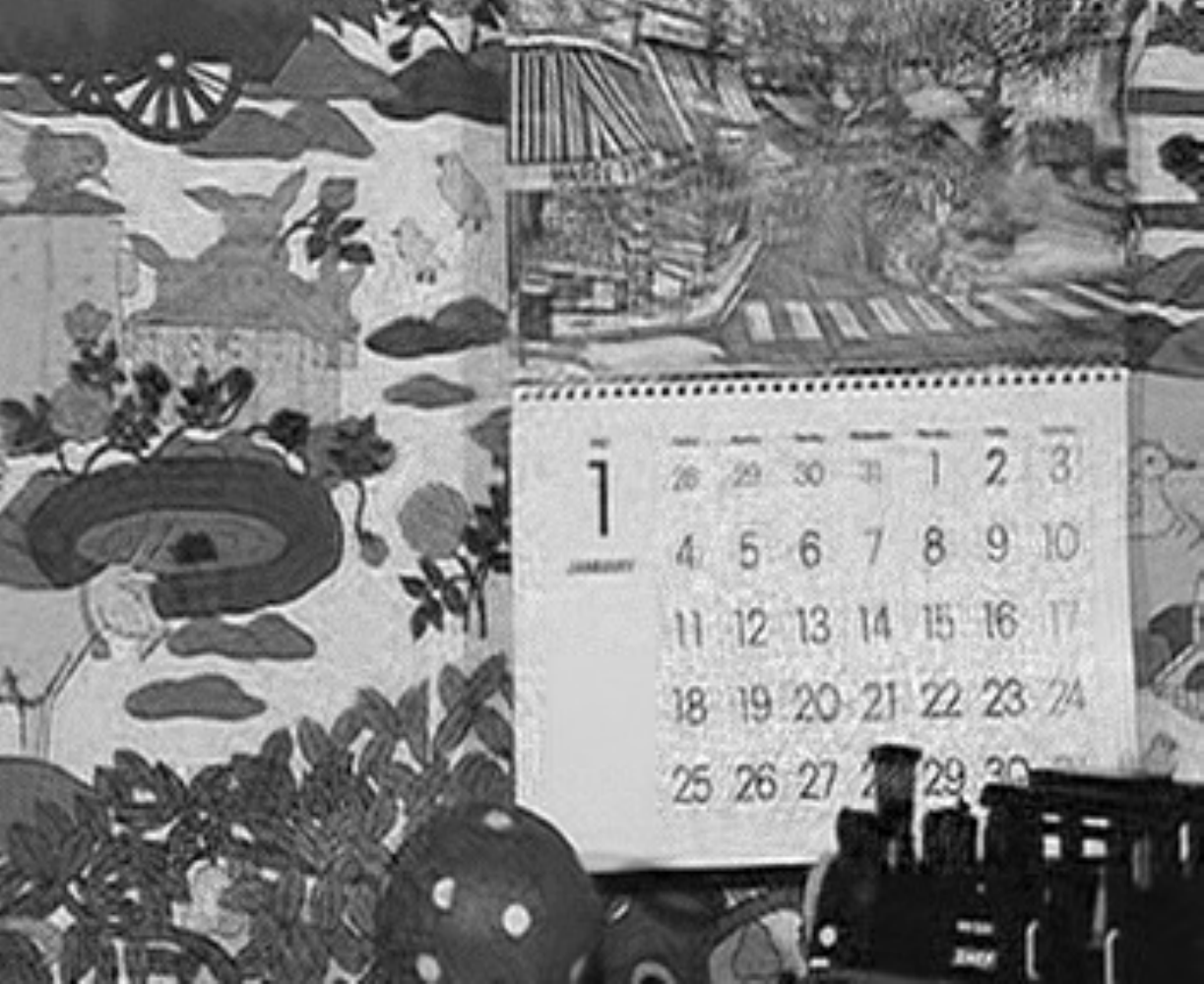}
                 \centering
                 \caption{}
                 \label{fig:1}
         \end{subfigure}%
         \begin{subfigure}[b]{0.19\textwidth}
                 \includegraphics[scale=0.265]{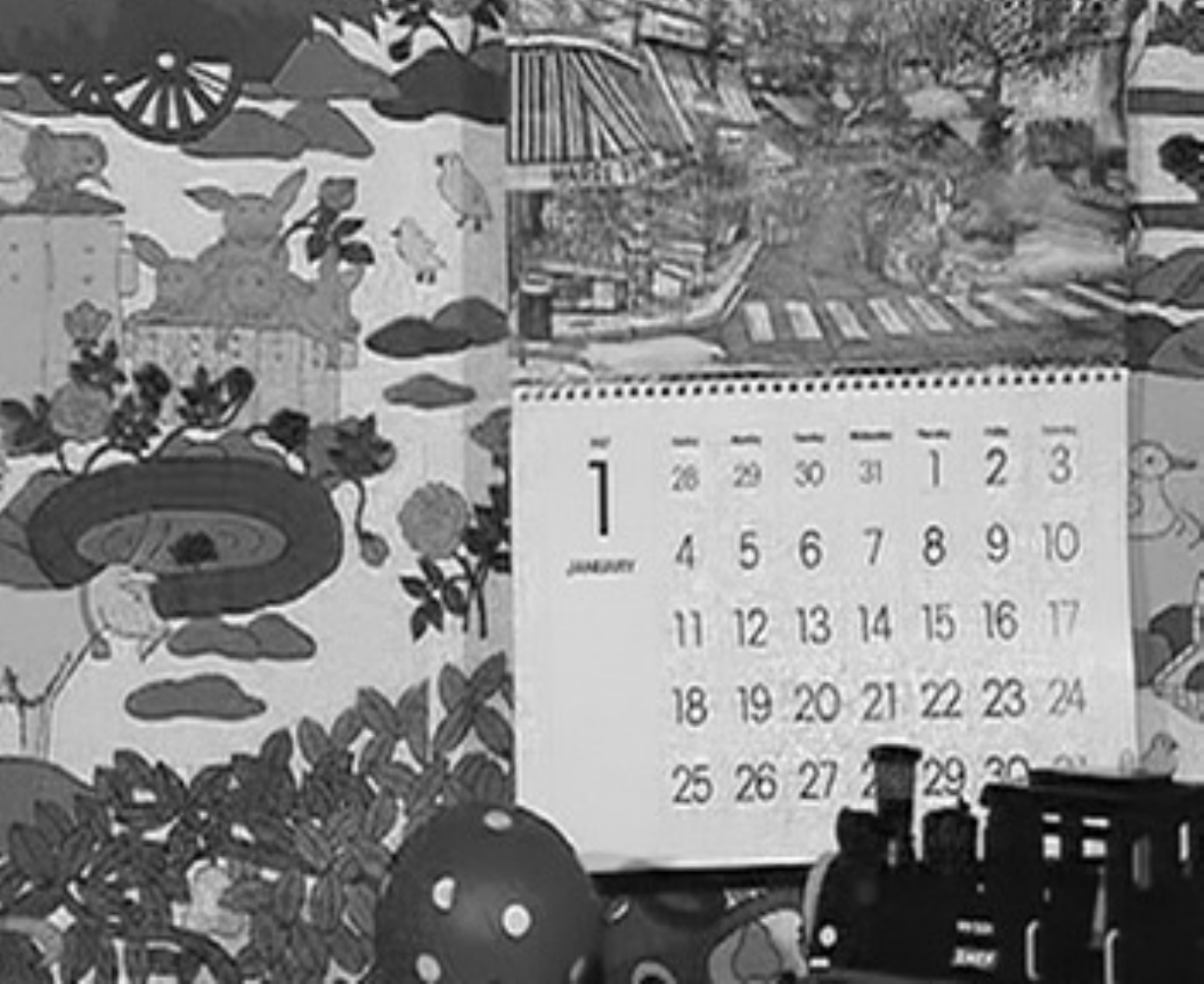}
                 \centering
                 \caption{}
                 \label{fig:1}
         \end{subfigure}\\
         \begin{subfigure}[b]{0.2\textwidth}
                 \includegraphics[scale=0.33]{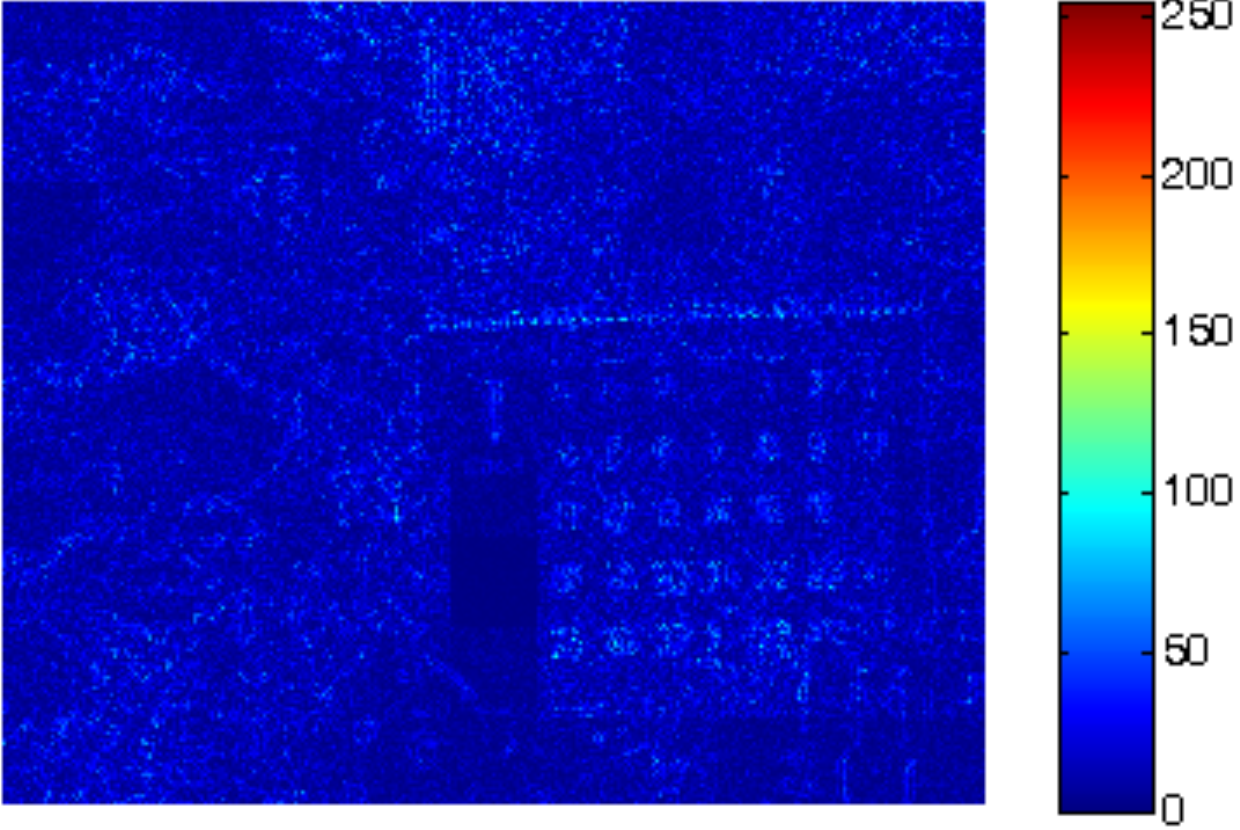}
                 \centering
                 \caption{}
                 \label{fig:1}
         \end{subfigure}
         \begin{subfigure}[b]{0.21\textwidth}
                 \includegraphics[scale=0.33]{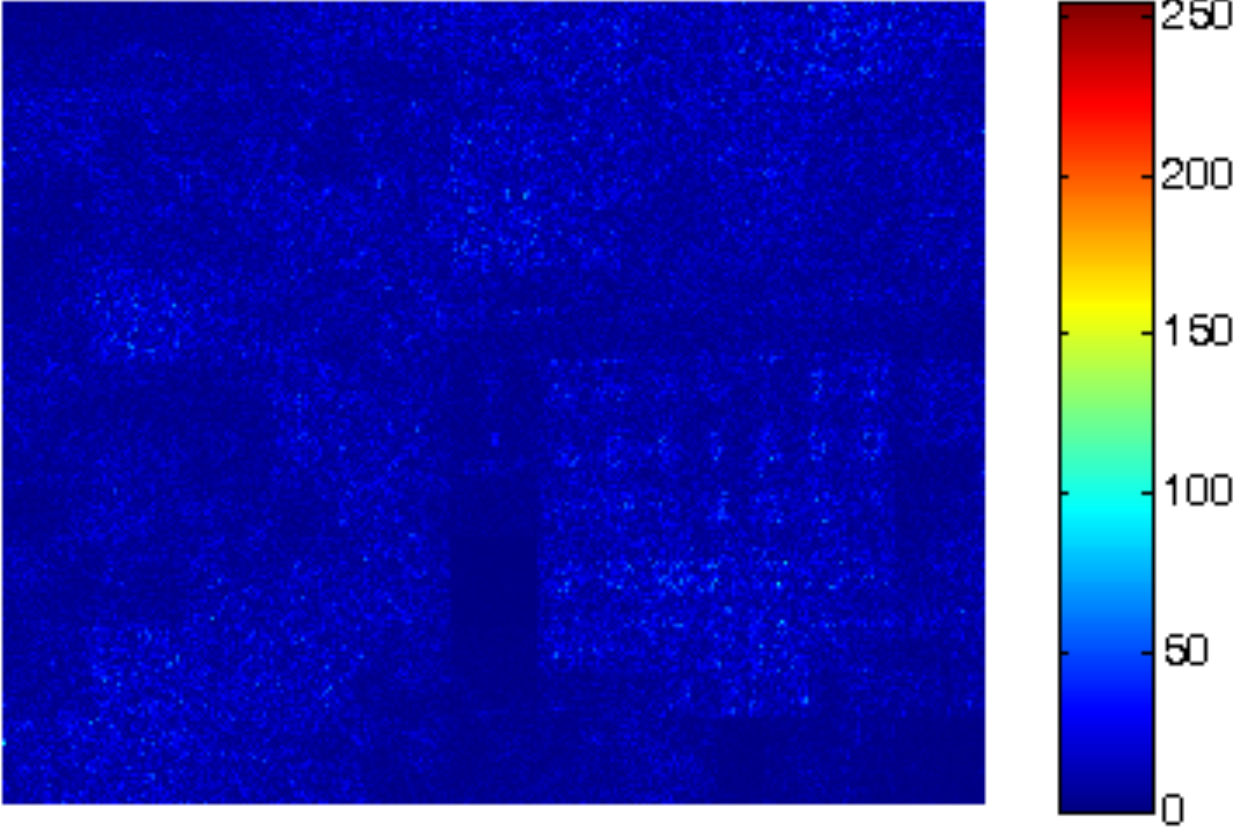}
                 \centering
                 \caption{}
                 \label{fig:1}
         \end{subfigure}%
         \begin{subfigure}[b]{0.2\textwidth}
                 \includegraphics[scale=0.33]{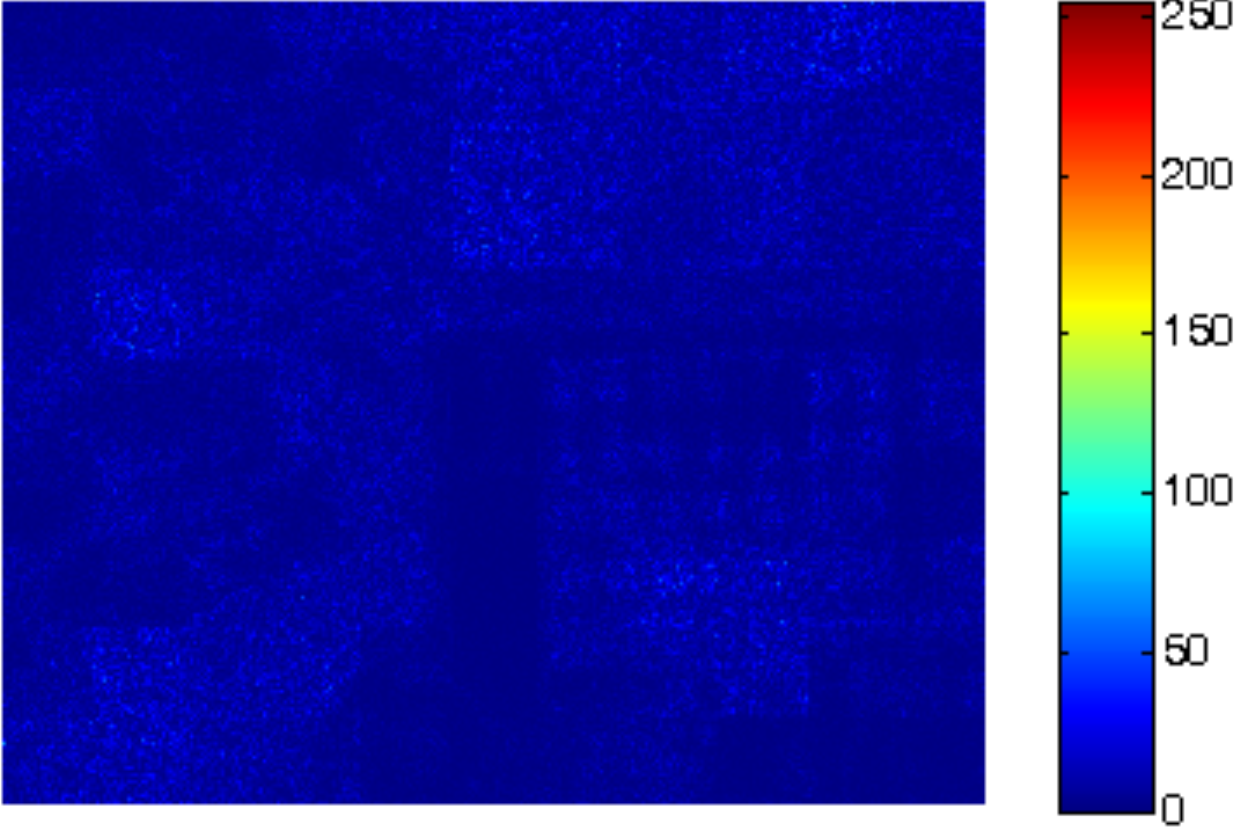}
                 \centering
                 \caption{}
                 \label{fig:1}
         \end{subfigure}\\}}
         {\footnotesize{Fig. 5: Different decoding of the 30\emph{th} frame of \emph{Mobile and Calendar} (w.r.t. recovery as a non-key frame $MR_{K}=0.4,MR_{NK}=0.4$). (a) Using the 2D DDWT basis intra-frame decoder (PSNR=22.02 dB, SSIM=0.645). (b) Using the MH method [18] (PSNR=24.1 dB, SSIM=0.749). (c) Using the proposed method (PSNR=26.6 dB, SSIM=0.863). (d)-(f) Magnitude of error reconstruction for (a)-(c), respectively.}}
\end{figure}
\begin{table}[!t]
\renewcommand{\arraystretch}{1}
{\footnotesize{\footnotesize{TABLE III:~\sc{ Performance Comparision in PSNR} \emph{($dB$)}  \sc and SSIM Values by Different Methods for The \emph{Mobile and Calendar} and The \emph{Hall Monitor} Video Sequences}}}\\
\label{table_example}
\centering
\begin{tabular}{c|c|c|c|c|c|}\cline{2-6}
&\multicolumn{5}{c|}{non-key frame measurement ratio (MR$_{NK}$):}\\ \cline{2-6}
 &0.1 & 0.2 & 0.3& 0.4 & 0.5\\ \hline 
\multicolumn{1}{|c|}{Algorithm} & \multicolumn{5}{c|}{Mobile and Calendar (CIF) MR$_K$=MR$_{NK}$}\\ \hline
\multicolumn{1}{|c|}{intra-frame 2D DDWT}&17.570&19.389&20.687&21.948&23.274\\
\multicolumn{1}{|c|}{}&0.383&0.495&0.575&0.644&0.703\\
\multicolumn{1}{|c|}{inter-frame MH [18]}&17.548&20.359&22.243&23.9754&26.047\\
\multicolumn{1}{|c|}{}&0.389&0.567&0.668&0.745&0.814\\ 
\multicolumn{1}{|c|}{Proposed method}&17.877&21.181&23.812&26.461&29.463\\
\multicolumn{1}{|c|}{}&0.435&0.647&0.772&0.862&0.924\\\hline\hline
\multicolumn{1}{|c|}{Algorithm} & \multicolumn{5}{c|}{Mobile and Calendar (CIF) MR$_K$=0.5}\\ \hline
\multicolumn{1}{|c|}{intra-frame 2D DDWT}&18.727&20.189&21.214&22.217&23.274\\
\multicolumn{1}{|c|}{}&0.448&0.536&0.600&0.655&0.703\\
\multicolumn{1}{|c|}{inter-frame MH [18]}&19.255&21.518&23.001&24.355&26.047\\
\multicolumn{1}{|c|}{}&0.475&0.618&0.698&0.759&0.814\\ 
\multicolumn{1}{|c|}{Proposed method}&20.180&22.875&24.938&27.009&29.463\\
\multicolumn{1}{|c|}{}&0.534&0.708&0.807&0.879&0.924\\\hline\hline
\multicolumn{1}{|c|}{Algorithm} & \multicolumn{5}{c|}{Halll Monitor (CIF) MR$_K$=MR$_{NK}$}\\ \hline
\multicolumn{1}{|c|}{intra-frame 2D DDWT}&23.528&28.893&32.127&34.722&37.082\\
\multicolumn{1}{|c|}{}&0.761&0.888&0.927&0.949&0.963\\
\multicolumn{1}{|c|}{inter-frame MH [18]}&25.291&29.608&32.597&34.953&37.209\\
\multicolumn{1}{|c|}{}&0.814&0.896&0.930&0.947&0.961\\ 
\multicolumn{1}{|c|}{Proposed method}&26.133&31.428&35.414&38.765&41.379\\
\multicolumn{1}{|c|}{}&0.845&0.926&0.953&0.970&0.979\\\hline\hline
\multicolumn{1}{|c|}{Algorithm} & \multicolumn{5}{c|}{Hall Monitor (CIF) MR$_K$=0.5}\\ \hline
\multicolumn{1}{|c|}{intra-frame 2D DDWT}&26.235&30.530&33.109&35.194&37.082\\
\multicolumn{1}{|c|}{}&0.849&0.919&0.934&0.952&0.963\\
\multicolumn{1}{|c|}{inter-frame MH [18]}&27.525&31.003&33.419&35.304&37.209\\
\multicolumn{1}{|c|}{}&0.867&0.921&0.936&0.952&0.961\\ 
\multicolumn{1}{|c|}{Proposed method}&28.983&33.268&36.447&39.189&41.379\\
\multicolumn{1}{|c|}{}&0.890&0.935&0.959&0.972&0.979\\\hline
\end{tabular}
\end{table}
\section{Conclusion}
The motivation of this paper is to propose a novel framework for CVS. In our proposed method, we employed the ALS basis and $\ell_0$ minimization for recovering of key frames, while non-key frames are recovered by firstly initialing a prediction of current non-key frame using previous reconstructed frame (in order to exploit the temporal redundancy), and then adopting the prediction into a proper optimization problem. Also, we investigated the effectiveness of three well-known dictionary learning algorithms (in order to learning an ALS basis), and we found out that MDU provides a better recovery performance (in quality) compared to the K-SVD and MOD, but at the cost of higher computational complexity. We found it reasonable to use the K-SVD as the dictionary learning algorithm in our proposed scheme. The numerical results show the adequacy of our proposed method in CVS, compared to the mentioned methods.

\section*{Appendix I}
Consider the following minimization of the quadratic function:
$$\min_{v}{Q_{1}(v)}=\min_{v}{{1\over 2}{\|f-\Phi v\|_{\ell_2}^2}+{\frac{\mu}{2}}{\|D\alpha^{k}-v-b^{k}\|_{\ell_{2}}^2}}, \eqno(38)$$
that can be solved by setting its gradient to be zero:
$$\nabla Q_{1}={g}_{1}^{k}=-\Phi^{T}(f-\Phi v)-\mu (D\alpha^{k}-v-b^{k})$$
$$=-\Phi^{T}f+\Phi^{T}\Phi v-\mu(D\alpha^{k}-b^{k})+\mu v=0$$
$$\Rightarrow(\Phi^{T}\Phi+\mu I)v=\mu(D\alpha^{k}-b^{k})+\Phi^{T}f$$
$$\Rightarrow v=v^{k+1}=(\Phi^{T}\Phi+\mu I)^{-1}\big(\mu(D\alpha^{k}-b^{k})+\Phi^{T}f\big).$$
In order to avoid using matrix inverse, the gradient descent method is utilized, i.e., 
$$v^{k+1}=v^{k}+\eta_{1}^{k} g_{1}^{k}, ~~\eta_{1}^{k}>0. \eqno(39)$$
By incorporating (39) into (38), and setting its partial derivative with respect to $\eta_{1}^{k}$ to be zero, the optimal step size, $\eta_{1}^{k}$, is yielded, i.e., (for simplicity, the subscript $k$ in $\eta_{1}^{k}$ and $g_{1}^{k}$ is omitted without confusion)
$${{\partial Q_{1}}\over {\partial \eta_{1}}}=g_{1}^{T}\Phi^{T}(f-\Phi v+\Phi\eta_{1} g_{1})+\mu g_{1}^{T}(D\alpha^{k}-v+\eta_{1} g_{1}-b^{k})=0,$$
$$\Rightarrow g_{1}^{T}\Phi^{T}\Phi \eta_{1} g_{1}+\mu g_{1}^{T}\eta_{1} g_{1}=-g_{1}^{T}\Phi^{T}(f-\Phi v)-\mu g_{1}^{T}(D\alpha^{k}-v-b^{k}),$$
$$\Rightarrow g_{1}^{T}(\Phi^{T}\Phi+\mu I)\eta_{1} g_{1}=g_{1}^{T}\big(\underbrace{-\Phi^{T}(f-\Phi v)-\mu (D\alpha^{k}-v-b^{k})}_{g_{1}}\big)$$
$$\Rightarrow \eta_{1}={\rm abs}\big({{g_{1}^{T}g_{1}}\over{g_{1}^{T}(\Phi^{T}\Phi+\mu I) g_{1}}}\big).$$
\section*{Appendix II}
The objective function in (33) is given by
$$f(\alpha,\alpha_{{t-1}_{l}})={1 \over 2}\|D\alpha_{l}-r_{p_{l}}\|_{\ell_2}^{2}+{\theta_{1,l}}\|\alpha_l\|_{\ell_1}+{\theta_{2,l}}\|\alpha_l-\alpha_{{t-1}_{l}}^{*}\|_{\ell_1},)$$
Let us add to it the following term
$${\rm dist}(\alpha_l,\alpha_0)={c \over 2}\|\alpha_{l}-{\alpha_{0}}\|_{\ell_2}^{2}-{1 \over 2}\|D\alpha_{l}-{D\alpha_{0}}\|_{\ell_2}^{2},$$
where the parameter $c$ is chosen in such a way that the function ${\rm dist}(\cdot,\cdot)$ is strictly convex (with respect to $\alpha_l$), implying that its Hessian should be positive-definite, $cI-D^{T}D>0$. This is satisfied by the choice $c>\|D^{T}D\|_{\ell_2}=\lambda_{\max}(D^{T}D)$ (the maximal eigenvalue of the matrix $D^{T}D$). The rationale behind this manipulation is to convert the objective function to a new one, for which we are able to get a closed-form expression for its global minimizer. The new objective function in (40) is called a surrogate function
$$\tilde{f}(\alpha,\alpha_{{t-1}_{l}},\alpha_0)={1 \over 2}\|D\alpha_{l}-r_{p_{l}}\|_{\ell_2}^{2}+{\theta_{1,l}}\|\alpha_l\|_{\ell_1}+{\theta_{2,l}}\|\alpha_l-\alpha_{{t-1}_{l}}^{*}\|_{\ell_1}+{c \over 2}\|\alpha_{l}-{\alpha_{0}}\|_{\ell_2}^{2}-{1 \over 2}\|D\alpha_{l}-{D\alpha_{0}}\|_{\ell_2}^{2}. \eqno(40)$$
The above function can be simplified as follow
$$\tilde{f}(\alpha,\alpha_{{t-1}_{l}},\alpha_0)={1 \over 2}\|D\alpha_{l}\|_{\ell_2}^{2}+{1 \over 2}\|r_{p_{l}}\|_{\ell_2}^{2}-\alpha_{l}^{T}D^{T}r_{p_{l}}+{\theta_{1,l}}\|\alpha_l\|_{\ell_1}$$$$+{\theta_{2,l}}\|\alpha_l-\alpha_{{t-1}_{l}}^{*}\|_{\ell_1}+{c \over 2}\|\alpha_{l}\|_{\ell_2}^{2}+{c \over 2}\|\alpha_{0}\|_{\ell_2}^{2}-c\alpha_{l}^{T}\alpha_{0}$$$$-{1 \over 2}\|D\alpha_{l}\|_{\ell_2}^{2}-{1 \over 2}\|D\alpha_{0}\|_{\ell_2}^{2}+\alpha_{l}^{T}D^{T}D\alpha_{0}$$
and by doing more simplification, it yields
$$\tilde{f}(\alpha,\alpha_{{t-1}_{l}},\alpha_0)={\rm const_{1}}+{\theta_{1,l}}\|\alpha_l\|_{\ell_1}+{\theta_{2,l}}\|\alpha_l-\alpha_{{t-1}_{l}}^{*}\|_{\ell_1}+{c \over 2}\|\alpha_{l}\|_{\ell_2}^{2}-c \alpha_{l}^{T} {\rm v_{0}}, \eqno(41)$$
where ${\rm v_{0}}={1 \over c}D^{T}(r_{p_{l}}-D\alpha_{0})+\alpha_{0}$.\\
As can be seen, the term $\|D\alpha_{l}\|_{\ell_2}^{2}$ drops in the new function that makes the minimization much simpler. Also, the term $\rm const_1$ in (41) implying all the terms that are dependent on $r_{p_{l}}$ and $\alpha_{0}$ alone.

By simplifying and reorganizing (41), the surrogate objective function can be written as
$$\tilde{f}(\alpha,\alpha_{{t-1}_{l}},\alpha_0)={\rm const_{2}}+{{\theta_{1,l}}\over c}\|\alpha_l\|_{\ell_1}+{{\theta_{2,l}}\over c}\|\alpha_l-\alpha_{{t-1}_{l}}^{*}\|_{\ell_1}+{1 \over 2}\|\alpha_{l}-\alpha_{0}\|_{\ell_2}^{2}.$$
The minimizer of the above function can be obtained by an iterative-shrinkage procedure, producing sequence of results ${\{{\alpha_{{{l}}_i}^{{{k}^\prime}}}\}}_i$ (the subscript $i$ denotes the $i$th entry in $\alpha_{{l}}$, and the subscript $k^{\prime}$ implies the iteration number), where at the $k+1$\emph{th} iteration, $\tilde{f}$ (and of course $f$) is minimized with the assignment $\alpha_0=\alpha_{l}^{k}$ as follow
$${{\alpha_{{{l}}_i}^{{{k}^\prime}+1}}}={\rm sgn}(\alpha_{{t+1}_{l_{i}}}^*) {\cal S}_{{\theta_{1}^{\prime}},{{\theta_{2}^{\prime}}},{\alpha_{{t+1}_{l_{i}}}^*}}\big({\rm sgn}(\alpha_{{t+1}_{l_{i}}}^*) {\rm v}_{i}^{k^{\prime}}\big),$$
where$${\theta_{\chi}^{\prime}}={{\theta_{\chi , l}}\over c}, \chi=\{1,2\},$$
$${\rm v^{k^{\prime}}}={1 \over c}D^{T}(r_{p_{l}}-D\alpha_{{l}}^{k^{\prime}})+\alpha_{{l}}^{k^{\prime}},$$
and the generalized shrinkage operator defines as follow$${\cal S}_{{\theta_{1}^{\prime}},{{\theta_{2}^{\prime}}},{\rho}}(x)=\left \{\begin{array}{cc} x+{{\theta_{1}^{\prime}}}+{{\theta_{2}^{\prime}}}, & x<-{{\theta_{1}^{\prime}}}-{{\theta_{2}^{\prime}}}\\0, & -{{\theta_{1}^{\prime}}}-{{\theta_{2}^{\prime}}}\leq x\leq {{\theta_{1}^{\prime}}}-{{\theta_{2}^{\prime}}}\\x-{{\theta_{1}^{\prime}}}+{{\theta_{2}^{\prime}}},&{{\theta_{1}^{\prime}}}
-{{\theta_{2}^{\prime}}}<x<{{\theta_{1}^{\prime}}}-{{\theta_{2}^{\prime}}}+{{\rho}}\\{{\rho}}
,&{{\theta_{1}^{\prime}}}-{{\theta_{2}^{\prime}}}+{{\rho}}\leq x \leq {{\theta_{1}^{\prime}}}+{{\theta_{2}^{\prime}}}+{{\rho}}\\
x-{{\theta_{1}^{\prime}}}-{{\theta_{2}^{\prime}}} & x>{{\theta_{1}^{\prime}}}+{{\theta_{2}^{\prime}}}+{{\rho}} \\ \end{array}\right.$$

\section*{Acknowledgments}
The authors would like to thank Prof. W. Dong (Xidian University) and Dr. J. Zhang (Peking University) for many fruitful discussions. They also would like to express their gratitude to the authors of [18], [26] and [42] for sharing the source code of their papers used in Section IV.

\ifCLASSOPTIONcaptionsoff
  \newpage
\fi



%

{\emph{This paper was presented in part in 7th International Symposium on Telecommunications, Tehran, Iran, Sep. 2014.}}\\
{\footnotesize{N. Eslahi, A. Aghagolzadeh, and S. M. H. Andargoli, ``Recovery of compressive video sensing via dictionary learning and forward prediction," in \emph{Proc. 7th Int. Symp. Telecommun. (IST)}, Sep. 2014, pp. 833-838.}}

\end{document}